\documentclass{article}
\usepackage{amssymb}
\usepackage{amsfonts}
\usepackage{amsmath}
\usepackage[left]{lineno}

\input{tcilatex}
\begin{document}

\author{J. G. Cardoso\thanks{%
jorge.cardoso@udesc.br}\thanks{%
Retired from the Santa Catarina State University.} \\
Department of Mathematics\\
Centre for Technological Sciences-UDESC\\
89223-100 Joinville Santa Catarina\\
Brazil.}
\title{A Pseudo-Unitary Version of Schwinger's Symbolism of Atomic
Measurements and a Prospect for a New Relativistic Quantum Information Theory%
}
\date{ }
\maketitle

\begin{abstract}
We transcribe into a relativistic dynamical framework proposed earlier the
measurement processes that are traditionally described within the realm of
non-relativistic quantum mechanics. The correlations between observations
registered in different spacetime frames, concerning intermediate steps and
outcomes of microscopic measurements, are attained through the
implementation of the orthochronous proper Poincar\'{e} subgroup of an
appropriate realization of $SU(2,2)$. Such correlations turn out to support
the view whereby the physical inner structure of the Schr\"{o}dinger quantum
mechanical picture may be formally described within special relativity. It
appears that the overall work affords a fundamental background to the
construction of covariant quantum computational gates whilst making it
feasible to elaborate upon the observed formation of entangled states in
processes of creation-annihilation pairs. The important question as to
whether quantum computation should bear an invariant character is then
raised.
\end{abstract}

\section*{Introduction}

The leading assumption upon which the construction of the classical theory
of measurement rests, lies in the fact that the interaction between the
measuring devices and the systems being observed can always be made
arbitrarily small. It follows that every classical measurement process may
be carried out without disturbing significantly any physical attribute of
the system involved in the process. On the other hand, the interaction
between measurement instruments and atomic systems produces disturbances on
the systems that can not be fully controlled and predicted [1-7], whence any
such disturbance can not be made arbitrarily small or even precisely
compensated by any means. Any process aimed at measuring an observable
attribute of some atomic system thus holds up indivisibility in the sense
that any attempt to trace the history of the system during the measurement
would entail modifying the intrinsic characteristics of the process, while
likewise producing unavoidable changes in previously known values for other
attributes of the system [2]. It is this property of microscopic
measurements which typically precludes dynamical variables of atomic systems
from being represented by classical numbers.\footnote{%
The most remarkable application of the theory of quantum measurements has
thus far been addressed to a non-relativistic approach often designated in
the literature as measurement-based quantum computation (see, for instance,
Refs. [8-12]).}

In the present paper, the symbolism of atomic measurements that had been
developed by Schwinger a long ago [2,3] towards describing the traditional
class of non-relativistic quantum measurement processes, is transcribed into
the dynamical framework built up in Ref. [13]. The key structure upon which
the construction of this framework stands amounts to Cartan's space, the
four-valued representation space of the restricted conformal group for
special relativity [14]. A fundamental physical character was ascribed to
Cartan's space thereabout, and a supposed relativistic quantum measurement
process was thus described within a finite-dimensional framework wherein the
dynamics of free particles and antiparticles could seemingly be formulated
in a purely quantum mechanical way.

What has supplied the underlying basis for the ascription of an elementary
character to Cartan's space, stems primarily from the existence within the
theoretical structure of special relativity of a complex indefinite inner
product space endowed with a representative group of pseudo-unitary linear
transformations, which could thereby bring forward in a non-trivial way the
essential role that is usually played by complex numbers in any quantum
mechanical formulation. A fairly complete algebraic description of Cartan's
space had been given earlier in Refs. [15,16] where a detailed covariant
account is provided of $SU(2,2)$, the group of unimodular pseudo-unitary $%
(4\times 4)$-matrices which effectively acts on any copy of Cartan's space
as a fifteen-parameter group of linear transformations that leave invariant
quadratic forms of signature $(++--)$. It then became apparent that the
overall framework could supply a relativistic quantum mechanical description
of free twofold systems.

Remarkably enough, the pseudo-unitary framework of Ref. [13] was thought of
as greatly unifying the description of free particles and antiparticles of
any spin, whence it has appeared to differ profoundly from the famous van
der Waerden two-component formulation of Dirac's equation which at first
describes the helicity degrees of freedom of one and the same spin-$1/2$
particle. The crude absence of unitarity from Dirac's theory thus seemed to
have been bypassed inside a manifestly mechanical framework. Just based upon
Naimark's theorems [17,18], which ensure the existence in Cartan's space of
common eigenvectors for commuting linear operators, every particle was
physically characterized through a complete set of compatible observables.
Spaces of state vectors for any free particles and antiparticles were
accordingly taken as two-complex-dimensional subspaces that occur in
orthogonal decompositions of isomorphic copies of Cartan's space, whereas
all dynamical variables were taken to carry a symbolic coordinate-free
character having twofold discrete spectra, and thence to operate linearly on
spaces of states independently of the action of any generator of the
background restricted conformal group. Each of these variables shows up as
an operator restriction that has one of the two pieces borne by the
orthogonal decomposition of a copy of Cartan's space as its principal
invariant eigenspace, with the other piece being considered as the zero
subspace of the variable in question. Orbital angular momenta were assumed
to be absent from dynamical sets such that only spin-helicity and
polarization degrees of freedom should be accounted for as regards the
arrangements of angular-momentum contributions to eventual preparations of
states. Thus, electric and all the other flavour-colour charges, spins,
helicities, polarizations as well as total energies, were the only
quantities that should be taken to constitute any complete sets of
observables.

We will see that a set of symbolic correlations, which directly arises out
of the implementation of the operationality of an adequate realization of $%
SU(2,2)$, turns out to support the contextual view whereby the physical
inner structure of the Schr\"{o}dinger quantum mechanical picture can be
formally described within the observational framework of special relativity,
while maintaining the concept of Schr\"{o}dinger's quantum mechanical states
along with the old postulated interpretation of Born's probabilities. The
work to be exhibited herein may presumptively enable an implementation in
spacetime frames of several measurement processes on a particular class of
physical systems, whose outcomes are subjected to clearly defined
observational correlations. We believe that the work bears an inherent
physical significance in that it may pave the way for the construction of a
fundamental background which could allow systematic settlements of covariant
quantum computational gates to be obtained.

Indeed, we will consider only the mechanical situation of composite
ensembles constituted by pairs of companion particles and antiparticles that
can be measured separately. All dynamical spectra shall have neither
geometric nor algebraic degeneracy. As emphasized in Ref. [2], the
usefulness of measurement processes generally requires the presence of
incompatible observables during the occurrences of the processes, but we
will avoid this limitation by implementing measurement operators that
suitably absorb amplitudes of eigenstates for compatible observables whilst
supplying covariantly non-trivial statistical relations between
intermediate-stage and outgoing states. It appears that the creation and
destruction of states eventually caused by local measurements in the absence
of spectral degeneracies, can be made into invariant properties of
relativistic observations while Born probabilities can not.

It shall be necessary for us to call upon some of the configurations
displayed in Refs. [13,15,16] before going into the dynamics of measurement
processes. Some review material concerning Cartan's space as well as the
algebraic structure of $SU(2,2)$ and traces of operator restrictions in
Cartan's space, is thus brought in Sections 2 through 4. In Section 5, we
define prototypes for states, observables and local spectra prior to our
transcription. All the corresponding observational correlations are shown in
Section 6. In Section 7, we will draw up an outlook related to some
important quantum mechanical issues and to our belief in the physical
significance of the work. The terminology borne by the theory of linear
operators in vector spaces equipped with indefinite inner products [19-23],
will be partially used throughout the work. For the sake of convenience, we
will make use of the "reflected" bra-ket and index notations explained in
Section 1 of Ref. [13]. Particularly, bra-kets carrying double or single
angular brackets shall denote Hilbert or indefinite inner products,
respectively. In case an operator occurs in a bra-ket product, the vector on
which it acts will bear a single bar and the specification of its action
will be made up by attaching a double bar to the other vector. When there
are operators acting on both sides of a bra-ket product, their action will
be stipulated by inserting a single bar between the apposite operator
blocks. Basis vectors, measurement operators and created states will be
labelled by lower-case Greek letters surrounded by round brackets, while the
labels for any vector components, matrix entries and eigenvalues will be
taken as non-bracketed lower-case Greek letters. Some further conventions
will be explained occasionally. The works of Refs. [13,15,16] shall be used
many times in Sections 2 and 3, but they will henceforth be referred to
explicitly just a few times.

\section*{Cartan's space}

We initially consider the Hilbert space $\mathcal{H}=\left( \mathbb{C}^{4},%
\mathcal{D}_{I}\right) $, with $\mathcal{D}_{I}$ being the usual
positive-definite inner product on $\mathbb{C}^{4}$. In the canonical basis $%
\left\{ <e_{(\mu )}\mid \right\} $ for $\mathbb{C}^{4}$, we write the dual
linear combinations for some $<x\mid \in \mathbb{C}^{4}$%
\begin{equation}
<x\mid =x^{\mu }<e_{(\mu )}\mid ,\quad \mid x>=\mid e_{(\nu )}>\overline{%
x^{\nu }},  \label{1}
\end{equation}%
such that\footnote{%
Here as elsewhere we have adopted the summation convention. A horizontal bar
lying over a symbol should denote the ordinary operation of complex
conjugation.}%
\begin{equation}
\mathcal{D}_{I}(<x\mid ,<y\mid )=\ll x\mid y\gg =x^{\mu }\Delta _{\mu \nu }%
\overline{y^{\nu }},  \label{2}
\end{equation}%
with $\left( \Delta _{\mu \nu }\right) $ denoting the identity $(4\times 4)$%
-matrix whose entries are expressed as%
\begin{equation}
\Delta _{\mu \nu }=\ll e_{(\mu )}\mid e_{(\nu )}\gg .  \label{3}
\end{equation}

Cartan's space is defined as the pair $\mathfrak{C}=\left( \mathbb{C}^{4},%
\mathcal{I}_{g}\right) $. In this definition, $\mathcal{I}_{g}$ stands for
an indefinite inner product on $\mathbb{C}^{4}$ which is specified by a
completely invertible linear operator $g$ according to the prescription%
\begin{equation}
g:\left( x^{0},x^{1},x^{2},x^{3}\right) \mapsto \left(
x^{0},x^{1},-x^{2},-x^{3}\right) .  \label{4}
\end{equation}%
Thus, $g=g^{-1}$ throughout $\mathbb{C}^{4}$ whereas%
\begin{equation}
\mathcal{I}_{g}(<x\mid ,<y\mid )=<x\mid y>_{g}=x^{\mu }g_{\mu \nu }\overline{%
y^{\nu }},  \label{5}
\end{equation}%
with the representation\footnote{%
The symbols $0_{2}$ and $I_{2}$ denote the zero and identity $(2\times 2)$%
-matrices.}%
\begin{equation}
g_{\mu \nu }=<e_{(\mu )}\mid e_{(\nu )}>_{g},\text{ }(g_{\mu \nu })=\left( 
\begin{array}{cc}
I_{2} & 0_{2} \\ 
0_{2} & -I_{2}%
\end{array}%
\right)  \label{6}
\end{equation}%
and the inner-product relationship%
\begin{equation}
<x\mid y>_{g}=\ll x\mid g\parallel y\gg .  \label{7}
\end{equation}

The operator $g$ bears Hermiticity and unitarity with respect to $\mathcal{D}%
_{I}$ as well as pseudo-Hermiticity and pseudo-unitarity with respect to $%
\mathcal{I}_{g}$. An invertible linear operator $A$ in $\mathfrak{C}$ is
said to be pseudo-Hermitian iff $A=A^{\bigstar }$, in conjunction with the
definition%
\begin{equation}
<x\mid A\parallel y>_{g}=<x\parallel A^{\bigstar }\mid y>_{g}.  \label{8}
\end{equation}%
If $A$ is also pseudo-unitary, we can write%
\begin{equation}
A^{-1}=A^{\bigstar }=A,  \label{9}
\end{equation}%
whence%
\begin{equation}
<x\mid A\mid A\mid y>_{g}=<x\parallel A^{\bigstar }A\mid y>_{g}=<x\mid
AA^{\bigstar }\parallel y>_{g}=<x\mid y>_{g}.  \label{10}
\end{equation}

One of the main geometric properties of $\mathfrak{C}$ concerns the
existence of pairs of $g$-orthogonal pseudo-Hermitian projectors in it
[19,21]. To any pair $\left( P^{+},P^{-}\right) $ of such projectors, there
corresponds a direct-sum splitting like%
\begin{equation}
\mathfrak{C}=\mathfrak{C}^{+}\oplus \mathfrak{C}^{-},  \label{11}
\end{equation}%
where\footnote{%
We have denoted the zero vector of $\mathfrak{C}$ as $<0\mid $. The
splitting (\ref{11}) is designated as a \textit{fundamental decomposition}
of $\mathfrak{C}$.}%
\begin{equation}
\mathfrak{C}^{\pm }=\left\{ <x\mid \in \mathfrak{C}\text{ such that }<x\mid
x>_{g}\gtrless 0\text{ or }<x\mid =<0\mid \right\} .  \label{12}
\end{equation}%
We thus have the vector splitting%
\begin{equation}
<x\mid =<x^{+}\mid +<x^{-}\mid ,  \label{13}
\end{equation}%
together with the property%
\begin{equation}
<x\mid P^{\pm }\parallel x>_{g}=<x\parallel P^{\pm }\mid x>_{g}  \label{14}
\end{equation}%
and the projections%
\begin{equation}
<e_{(\mu )}\mid P^{\pm }=<e_{(\mu )}^{\pm }\mid ,\text{ }<x\mid P^{\pm
}=<x^{\pm }\mid ,  \label{15}
\end{equation}%
such that%
\begin{equation}
<x^{+}\mid =\left( x^{0},x^{1},0,0\right) ,\text{ }<x^{-}\mid =\left(
0,0,x^{2},x^{3}\right) .  \label{16}
\end{equation}%
Hence, we can reexpress $\mathcal{I}_{g}$ as either of the prescriptions%
\begin{equation}
<x\mid y>_{g}=\ll x^{+}\mid y^{+}\gg -\ll x^{-}\mid y^{-}\gg  \label{17}
\end{equation}%
and%
\begin{equation}
g_{\mu \nu }=\ll e_{(\mu )}^{+}\mid e_{(\nu )}^{+}\gg -\ll e_{(\mu
)}^{-}\mid e_{(\nu )}^{-}\gg ,  \label{18}
\end{equation}%
while decomposing $A$ as%
\begin{equation}
<x\mid AP^{+}=<x^{+}\mid A^{++}+<x^{-}\mid A^{-+}  \label{19}
\end{equation}%
and%
\begin{equation}
<x\mid AP^{-}=<x^{+}\mid A^{+-}+<x^{-}\mid A^{--},  \label{20}
\end{equation}%
with the defining scheme%
\begin{equation}
\begin{array}{c}
A^{++}:\mathfrak{C}^{+}\rightarrow R(A)\cap \mathfrak{C}^{+},\text{ }A^{+-}:%
\mathfrak{C}^{+}\rightarrow R(A)\cap \mathfrak{C}^{-} \\ 
A^{-+}:\mathfrak{C}^{-}\rightarrow R(A)\cap \mathfrak{C}^{+},\text{ }A^{--}:%
\mathfrak{C}^{-}\rightarrow R(A)\cap \mathfrak{C}^{-},%
\end{array}
\label{21}
\end{equation}%
and $R(A)$ denoting the range of $A$. A formal decomposition of the matrix
entries%
\begin{equation}
A_{\mu \nu }=<e_{(\mu )}\mid A\parallel e_{(\nu )}>_{g}=A_{\mu }{}^{\lambda
}g_{\lambda \nu }  \label{22}
\end{equation}%
thus emerges out of implementing (\ref{21}). This procedure leads one, in
effect, to the following $(2\times 2)$-block representation for the operator 
$A$%
\begin{equation}
(A_{\mu \nu })=%
\begin{pmatrix}
<e_{(\mu )}^{+}\mid A^{++}\parallel e_{(\nu )}^{+}>_{g^{+}} & <e_{(\mu
)}^{+}\mid A^{+-}\parallel e_{(\nu )}^{-}>_{g^{-}} \\ 
<e_{(\mu )}^{-}\mid A^{-+}\parallel e_{(\nu )}^{+}>_{g^{+}} & <e_{(\mu
)}^{-}\mid A^{--}\parallel e_{(\nu )}^{-}>_{g^{-}}%
\end{pmatrix}%
.  \label{23}
\end{equation}

A \textit{restriction} $\func{Res}A_{\mathfrak{C}^{\pm }}$ of $A$ on $%
\mathfrak{C}^{\pm }$ is an invertible linear operator whose domain and range
are prescribed as [19]%
\begin{equation}
D\left( \func{Res}A_{\mathfrak{C}^{\pm }}\right) =\mathfrak{C}\cap \mathfrak{%
C}^{\pm },\text{ }R\left( \func{Res}A_{\mathfrak{C}^{\pm }}\right) \subset 
\mathfrak{C}^{\pm }\subset R(A).  \label{24}
\end{equation}%
When $A$ is restricted to $\mathfrak{C}^{\pm }$, we obtain the operator array%
\footnote{%
In Sections 5 and 6, the ranges of restrictions will be identified with $%
\mathfrak{C}^{\pm }.$}%
\begin{equation}
A=\left( 
\begin{array}{ll}
\func{Res}A_{\mathfrak{C}^{+}} & 0 \\ 
0 & \func{Res}A_{\mathfrak{C}^{-}}%
\end{array}%
\right) ,  \label{linn}
\end{equation}%
together with the expansion%
\begin{equation}
<x\mid A=<x^{+}\mid \func{Res}A_{\mathfrak{C}^{+}}+<x^{-}\mid \func{Res}A_{%
\mathfrak{C}^{-}}.  \label{29}
\end{equation}%
For operators $A,B$\ and $C$ in $\mathfrak{C},$ we can therefore write%
\begin{equation}
\func{Res}(ABC)_{\mathfrak{C}^{\pm }}=\func{Res}A_{\mathfrak{C}^{\pm }}\func{%
Res}B_{\mathfrak{C}^{\pm }}\func{Res}C_{\mathfrak{C}^{\pm }}.  \label{31}
\end{equation}

The restrictions on $\mathfrak{C}^{\pm }$ of the identity operator $I$ of $%
\mathfrak{C}$ lead to the representative entries%
\begin{equation}
<e_{(\mu )}^{\pm }\mid \func{Res}I_{\mathfrak{C}^{\pm }}\parallel e_{(\nu
)}^{\pm }>_{g^{\pm }}=g_{\mu \nu }^{\pm }=\pm <<e_{(\mu )}^{\pm }\mid
e_{(\nu )}^{\pm }>>,  \label{32}
\end{equation}%
which produce the reduced $(2\times 2)$-block matrices\footnote{%
We should stress that any $g$-restriction is allowed to be defined only on $%
\mathfrak{C}^{\pm }$.}%
\begin{equation}
\left( g_{\mu \nu }^{+}\right) =\left( 
\begin{array}{ll}
1 & 0 \\ 
0 & 1%
\end{array}%
\right) ,\text{ }\left( g_{\mu \nu }^{-}\right) =\left( 
\begin{array}{ll}
-1 & 0 \\ 
0 & -1%
\end{array}%
\right) ,  \label{33}
\end{equation}%
with $g^{\pm }=\func{Res}g_{\mathfrak{C}^{\pm }}$. It follows that, by
implementing (\ref{16}) together with the reduced identifications%
\begin{equation}
x^{0}=x_{+}^{0},\text{ }x^{1}=x_{+}^{1},\text{ }x^{2}=x_{-}^{0},\text{ }%
x^{3}=x_{-}^{1},  \label{34}
\end{equation}%
and setting the second of (\ref{15}) as%
\begin{equation}
<x^{\pm }\mid =x_{\pm }^{\mu }<e_{(\mu )}^{\pm }\mid ,  \label{35}
\end{equation}%
we recast the inner products on $\mathfrak{C}^{\pm }$ into the definite
configurations%
\begin{equation}
\mathcal{I}_{g^{\pm }}\left( <x^{\pm }\mid ,<y^{\pm }\mid \right) =<x^{\pm
}\mid y^{\pm }>_{g^{\pm }}=x_{\pm }^{\mu }g_{\mu \nu }^{\pm }\overline{%
y_{\pm }^{\nu }},  \label{36}
\end{equation}%
which agree with the pseudo-Hermiticity property%
\begin{equation}
<x^{\pm }\mid g^{\pm }\parallel y^{\pm }>_{g^{\pm }}=<x^{\pm }\parallel
g^{\pm }\mid y^{\pm }>_{g^{\pm }}.  \label{37}
\end{equation}%
Therefore, in agreement with (\ref{32}), one can also spell out the $%
\mathcal{I}_{g}$-expressions%
\begin{equation}
<e_{(\mu )}\mid P^{\pm }\parallel e_{(\nu )}>_{g}=<e_{(\mu )}^{\pm }\mid 
\func{Res}I_{\mathfrak{C}^{\pm }}\parallel e_{(\nu )}^{\pm }>_{g^{\pm }},
\label{38}
\end{equation}%
whereas the representation of $\func{Res}I_{\mathfrak{C}^{\pm }}$ with
respect to $\mathcal{D}_{I}$ is constituted by%
\begin{equation}
\ll e_{(\mu )}^{\pm }\mid \func{Res}I_{\mathfrak{C}^{\pm }}\parallel e_{(\nu
)}^{\pm }\gg =<<e_{(\mu )}\mid P^{\pm }\parallel e_{(\nu )}>>=\Delta _{\mu
\nu }^{\pm },  \label{39}
\end{equation}%
which right away yields the reduced matrices%
\begin{equation}
(\Delta _{\mu \nu }^{+})=\left( 
\begin{array}{ll}
1 & 0 \\ 
0 & 1%
\end{array}%
\right) =(\Delta _{\mu \nu }^{-}).  \label{40}
\end{equation}

The first adjoint $\mathfrak{C}^{\ast }=\left( \mathbb{C}^{4},\mathcal{I}%
_{g^{\ast }}\right) $ of $\mathfrak{C}$ is defined in such a way that each
element of $\mathfrak{C}$ enters a one-to-one linear mapping which produces
the basis relationships%
\begin{equation}
<e_{(\mu )}^{\pm }\mid \mapsto g_{\mu \lambda }^{\pm }<e_{\pm }^{\ast
(\lambda )}\mid ,\text{ }<e_{\pm }^{\ast (\mu )}\mid \mapsto g_{\pm }^{\ast
\mu \lambda }<e_{(\lambda )}^{\pm }\mid .  \label{41}
\end{equation}%
It should be evident that the operator rule for $g^{\ast }$ is formally the
same as the one for $g$ such that it is reasonable to bring forth the
splitting%
\begin{equation}
\mathfrak{C}^{\ast }=\mathfrak{C}_{+}^{\ast }\oplus \mathfrak{C}_{-}^{\ast },
\label{42}
\end{equation}%
with $\mathfrak{C}_{\pm }^{\ast }$ thus holding formally the same definition
as that given by (\ref{12}). Whence, for some element $<x^{\ast }\mid $ of $%
\mathfrak{C}^{\ast }$, we can take account of the component identifications%
\begin{equation}
x_{0}=x_{0}^{+},\text{ }x_{1}=x_{1}^{+},\text{ }x_{2}=x_{0}^{-},\text{ }%
x_{3}=x_{1}^{-}  \label{43}
\end{equation}%
for rewriting the adjoint linear combination%
\begin{equation}
<x^{\ast }\mid =x_{\mu }<e^{\ast (\mu )}\mid  \label{44}
\end{equation}%
in terms of the configurations%
\begin{equation}
<x_{\pm }^{\ast }\mid =x_{\mu }^{\pm }<e_{\pm }^{\ast (\mu )}\mid ,\text{ }%
x_{\mu }^{\pm }=x_{\pm }^{\lambda }g_{\lambda \mu }^{\pm },\text{ }x_{\pm
}^{\mu }=x_{\lambda }^{\pm }g_{\pm }^{\ast \lambda \mu }.  \label{45}
\end{equation}%
This procedure promptly yields%
\begin{equation}
<x^{\pm }\mid e_{(\mu )}^{\pm }>_{g^{\pm }}=x_{\mu }^{\pm },\text{ }<x_{\pm
}^{\ast }\mid e_{\pm }^{\ast (\mu )}>_{g_{\pm }^{\ast }}=x_{\pm }^{\mu },
\label{46}
\end{equation}%
and likewise brings out the reflexiveness property%
\begin{equation}
<x^{\pm }\mid x^{\pm }>_{g^{\pm }}=<x_{\pm }^{\ast }\mid x_{\pm }^{\ast
}>_{g_{\pm }^{\ast }}.  \label{linlin}
\end{equation}

Equations (\ref{41}) carry forward the canonical character of $<e_{(\mu
)}\mid $ to $<e^{\ast (\mu )}\mid $ such that the entries\footnote{%
It is clear that $\mathfrak{C}^{\pm }\simeq \mathbb{C}^{2}\simeq \mathfrak{C}%
_{\pm }^{\ast }$ such that the entry labels carried by any reduced or
restricted structures must assume the values 0 and 1.}%
\begin{equation}
g_{\pm }^{\ast \mu \nu }=<e_{\pm }^{\ast (\mu )}\mid e_{\pm }^{\ast (\nu
)}>_{g_{\pm }^{\ast }},\text{ }\Delta _{\pm }^{\ast \mu \nu }=\ll e_{\pm
}^{\ast (\mu )}\mid e_{\pm }^{\ast (\nu )}\gg ,  \label{lin}
\end{equation}%
coincide with those for $\func{Res}I_{\mathbb{C}^{\pm }}$. Additionally, the
entries (\ref{lin}) obey the relations\footnote{%
It is of some interest to remark that the $\ast $-adjointness obeys the same
rearrangement of factors as the one that occurs in pseudo-Hermitian and
Hermitian conjugations of matrix products. This will be used sometimes in
Section 6.}%
\begin{equation}
g_{\mu \nu }^{\pm }=\Delta _{\mu \lambda }^{\pm }g_{\pm }^{\ast \lambda
\sigma }\Delta _{\sigma \nu }^{\pm },\text{ }g_{\pm }^{\ast \mu \nu }=\Delta
_{\pm }^{\ast \mu \lambda }g_{\lambda \sigma }^{\pm }\Delta _{\pm }^{\ast
\sigma \nu },  \label{49}
\end{equation}%
along with the ones that are obtained from (\ref{49}) by just interchanging
the kernel letters $\Delta $ and $g$. For the restrictions carried by (\ref%
{32}), we thus have the decomposition%
\begin{equation}
\func{Res}I_{\mathfrak{C}^{\pm }}=\mid e_{(\mu )}^{\pm }>g_{\pm }^{\ast \mu
\nu }<e_{(\nu )}^{\pm }\mid ,  \label{50}
\end{equation}%
which shows us that $<e_{(\mu )}^{\pm }\mid $ possesses a completeness
property.

The unrestricted adjoint entries%
\begin{equation}
A^{\ast \mu \nu }=<e^{\ast (\mu )}\mid A^{\ast }\parallel e^{\ast (\nu
)}>_{g^{\ast }},  \label{51}
\end{equation}%
satisfy%
\begin{equation}
A_{\mu }{}^{\nu }=g_{\mu \lambda }A^{\ast \lambda \nu }\Longrightarrow
A_{\mu \nu }=g_{\mu \lambda }A^{\ast \lambda \sigma }g_{\sigma \nu }
\label{52}
\end{equation}%
and%
\begin{equation}
A^{\ast \mu }{}_{\nu }=g^{\ast \mu \lambda }A_{\lambda \nu }\Longrightarrow
A^{\ast \mu \nu }=g^{\ast \mu \lambda }A_{\lambda \sigma }g^{\ast \sigma \nu
}.  \label{53}
\end{equation}%
Hence, by adapting to (\ref{22}) the notation of (\ref{33}), likewise taking
up (\ref{51}), one obtains%
\begin{equation}
A_{\mu \nu }^{\pm }=<e_{(\mu )}^{\pm }\mid A^{\pm }\parallel e_{(\nu )}^{\pm
}>_{g^{\pm }},\text{ }A_{\pm }^{\ast \mu \nu }=<e_{\pm }^{\ast (\mu )}\mid
A_{\pm }^{\ast }\parallel e_{\pm }^{\ast (\nu )}>_{g_{\pm }^{\ast }}
\label{add1}
\end{equation}%
and%
\begin{equation}
A^{\pm }=\mid e_{(\mu )}^{\pm }>A_{\pm }^{\ast \mu \nu }<e_{(\nu )}^{\pm
}\mid ,\text{ }A_{\pm }^{\ast }=\mid e_{\pm }^{\ast (\mu )}>A_{\mu \nu
}^{\pm }<e_{\pm }^{\ast (\nu )}\mid ,  \label{54}
\end{equation}%
along with%
\begin{equation}
A_{\mu \nu }^{\pm }=g_{\mu \lambda }^{\pm }A_{\pm }^{\ast \lambda \sigma
}g_{\sigma \nu }^{\pm },\text{ }A_{\pm }^{\ast \mu \nu }=g_{\pm }^{\ast \mu
\lambda }A_{\lambda \sigma }^{\pm }g_{\pm }^{\ast \sigma \nu }  \label{55}
\end{equation}%
and%
\begin{equation}
\func{Res}I_{\mathfrak{C}_{\pm }^{\ast }}^{\ast }=\mid e_{\pm }^{\ast (\mu
)}>g_{\mu \nu }^{\pm }<e_{\pm }^{\ast (\nu )}\mid .  \label{56}
\end{equation}

The entire set of relationships between Hermitian and pseudo-Hermitian
conjugations in $\mathfrak{C},$ is supplied in Ref. [16]. We next recall
just the ones which may play a major role in Sections 5 and 6. A simpler
procedure for starting with involves using the restricted form of Eq. (\ref%
{7}), that is to say,%
\begin{equation}
<x^{\pm }\mid y^{\pm }>_{g^{\pm }}=\ll x^{\pm }\mid g^{\pm }\parallel y^{\pm
}\gg ,  \label{add900}
\end{equation}%
for pointing up that the definition%
\begin{equation}
<e_{(\mu )}^{\pm }\mid A^{\pm \bigstar }\parallel e_{(\nu )}^{\pm }>_{g^{\pm
}}=<e_{(\mu )}^{\pm }\parallel A^{\pm }\mid e_{(\nu )}^{\pm }>_{g^{\pm }}
\label{57}
\end{equation}%
amounts to the same thing as\footnote{%
From now onwards the "dagger" will be used to denote the Hermitian
conjugation for the Hilbert inner product.}%
\begin{equation}
\ll e_{(\mu )}^{\pm }\mid A^{\pm \bigstar }g^{\pm }\parallel e_{(\nu )}^{\pm
}\gg =\ll e_{(\mu )}^{\pm }\mid g^{\pm }\mid A^{\pm }\mid e_{(\nu )}^{\pm
}\gg =\ll e_{(\mu )}^{\pm }\mid g^{\pm }A^{\pm \dagger }\parallel e_{(\nu
)}^{\pm }\gg .  \label{58}
\end{equation}%
We are thus led to the operator relationship%
\begin{equation}
A^{\pm \bigstar }g^{\pm }=g^{\pm }A^{\pm \dagger }.  \label{59}
\end{equation}%
It follows that, setting\footnote{%
Henceforth the superscript $T$ will denote the usual operation of matrix
transposition.}%
\begin{equation}
A_{\mu \nu }^{\pm \bigstar }=<e_{(\mu )}^{\pm }\mid A^{\pm \bigstar
}\parallel e_{(\nu )}^{\pm }>_{g^{\pm }}=\overline{A_{\mu \nu }^{\pm T}}
\label{60}
\end{equation}%
and%
\begin{equation}
a_{\mu \nu }^{\pm \dagger }=\ll e_{(\mu )}^{\pm }\mid A^{\pm \dagger
}\parallel e_{(\nu )}^{\pm }\gg =\overline{a_{\mu \nu }^{\pm T}},  \label{61}
\end{equation}%
as well as making use of the formulae%
\begin{equation}
g_{\mu }^{\pm }{}^{\nu }=\Delta _{\mu \lambda }^{\pm }g_{\pm }^{\ast \lambda
\nu },\text{ }g_{\pm }^{\ast \mu }{}_{\nu }=\Delta _{\pm }^{\ast \mu \lambda
}g_{\lambda \nu }^{\pm }  \label{62}
\end{equation}%
and%
\begin{equation}
A_{\mu \nu }^{\pm \dagger }=<e_{(\mu )}^{\pm }\mid A^{\pm \dagger }\parallel
e_{(\nu )}^{\pm }>_{g^{\pm }}=a_{\mu \lambda }^{\pm \dagger }g_{\pm }^{\ast
\lambda }{}_{\nu },  \label{63}
\end{equation}%
yields the associations%
\begin{equation}
A_{\mu \nu }^{\pm \bigstar }=g_{\mu }^{\pm }{}^{\lambda }A_{\lambda \sigma
}^{\pm \dagger }g_{\pm }^{\ast \sigma }{}_{\nu }  \label{64}
\end{equation}%
and%
\begin{equation}
A_{\pm }^{\ast \bigstar \mu \nu }=g_{\pm }^{\ast \mu }{}_{\lambda }A_{\pm
}^{\ast \dagger \lambda \sigma }g_{\sigma }^{\pm }{}^{\nu },\text{ }A_{\pm
}^{\ast \dagger \mu \nu }=a_{\pm }^{\ast \dagger \mu \lambda }g_{\lambda
}^{\pm }{}^{\nu }.  \label{65}
\end{equation}

A pair of interesting relationships arises when we bring together (\ref{55}%
), (\ref{62}) and (\ref{64}). We have, in effect,%
\begin{equation}
A_{\mu \nu }^{\pm \bigstar }=\Delta _{\mu \lambda }^{\pm }A_{\pm }^{\ast
\dagger \lambda \sigma }\Delta _{\sigma \nu }^{\pm },\text{ }A_{\pm }^{\ast
\bigstar \mu \nu }=\Delta _{\pm }^{\ast \mu \lambda }A_{\lambda \sigma
}^{\pm \dagger }\Delta _{\pm }^{\ast \sigma \nu },  \label{66}
\end{equation}%
whence the entry arrays that constitute the matrices for each of the pairs $%
(A_{\mu \nu }^{\pm \bigstar },A_{\pm }^{\ast \dagger \lambda \sigma })$ and $%
(A_{\pm }^{\ast \bigstar \mu \nu },A_{\lambda \sigma }^{\pm \dagger }),$
coincide with one another.

\section*{The group SU$(2,2)$}

From the algebraic point of view, $SU(2,2)$ is the group constituted by the
usual operation of matrix multiplication and the set of unimodular matrices
that represent either of the classes $\left\{ \mathcal{U},\mathcal{U}^{\ast
}\right\} $ of pseudo-unitary operators in $\mathfrak{C}$ and $\mathfrak{C}%
^{\ast }$. The group $SU(2,2)$ appears as [23-29] as the \textit{special}
subgroup of $U(2,2)$ whose action leaves in a broader manner the definitions
(\ref{4})-(\ref{6}) invariant. We have%
\begin{equation}
SU(2,2)=\left\{ \breve{u}\in U(2,2)\text{ with }\func{det}\left( \breve{u}%
\right) =1\right\} ,  \label{67}
\end{equation}%
whence we can state that%
\begin{equation}
SU(2,2)=U(2,2)\cap SL(4,\mathbb{C})\subset SL(4,\mathbb{C}).  \label{68}
\end{equation}%
The maximal compact subgroup of $SU(2,2)$ amounts to [24]%
\begin{equation}
K=SU(2)\times U(1)\times SU(2)\subset SU(4),  \label{69}
\end{equation}%
where\footnote{%
In (\ref{70}), $U(n)$ denotes the group of unitary $(n\times n)$-matrices
for the Hilbert inner product, with $U(1)$ being particularly the
multiplicative group of unit-modulus complex numbers.}%
\begin{equation}
SU(n)=U(n)\cap SL(n,\mathbb{C}).  \label{70}
\end{equation}

An important property of $SU(2,2)$ comes from Cartan's decomposition theorem
[14,24,27], which asserts that there exists a one-to-one homeomorphism of $%
SU(2,2)$ onto the topological product%
\begin{equation}
T_{C}=(SU(2,2)\cap U(4))\times (SU(2,2)\cap P(4)),  \label{71}
\end{equation}%
where $P(4)$ is the group of positive-definite Hermitian $(4\times 4)$%
-matrices. This statement may be schematically displayed as%
\begin{equation*}
SU(2,2)\ni \breve{u}\longmapsto (U,H)\in T_{C},\text{ }U\in (SU(2,2)\cap
U(4)),\text{ }H\in (SU(2,2)\cap P(4)).
\end{equation*}%
Accordingly, we have the unique coupling prescription%
\begin{equation}
\breve{u}=UH,  \label{72}
\end{equation}%
such that we can likewise write%
\begin{equation}
\breve{u}^{\dagger }\breve{u}=H^{2},\text{ }\breve{u}\breve{u}^{\dagger
}=UH^{2}U^{-1}.  \label{73}
\end{equation}

It is shown in Ref. [15] that any adjoint realizations of $SU(2,2)$ are just
the same and possess closedness with respect to both pseudo-Hermitian and
Hermitian conjugations. In any case, the only admissible bases for
representing pseudo-unitary operators in $\mathfrak{C}$ and $\mathfrak{C}%
^{\ast }$ are $SU(2,2)$-related to each other.\footnote{%
The basis $\left\{ <e_{(\mu )}\mid \right\} $ is the so-called [13] \textit{%
computational basis} of $\mathfrak{C}$.} In what follows, we will reproduce
to some extent the presentation of the unstarred $g$-realization of $SU(2,2)$
as exhibited in Refs. [13,15]. If $u\in \mathcal{U}$, we may thus write the
configuration\footnote{%
The $g$-invariance exhibited by (\ref{74}) will be made explicit in Section
6.}%
\begin{equation}
uu^{\bigstar }=I\Leftrightarrow ugu^{\dagger }=g,  \label{74}
\end{equation}%
where $I$ denotes the identity operator of $\mathfrak{C}$ such as in Eq. (%
\ref{32}). Hence, utilizing a decomposition for each of $u$ and $u^{\bigstar
}$ of the same type as that given by (\ref{21}), yields the $\bigstar $%
-invariant operator relations%
\begin{equation}
u^{++}u^{++\bigstar }+u^{+-}u^{+-\bigstar }=I^{+},\text{ }%
u^{--}u^{--\bigstar }+u^{-+}u^{-+\bigstar }=I^{-}  \label{75}
\end{equation}%
and%
\begin{equation}
u^{++}u^{-+\bigstar }+u^{+-}u^{--\bigstar }=0=u^{-+}u^{++\bigstar
}+u^{--}u^{+-\bigstar },  \label{76}
\end{equation}%
where $I^{\pm }=\func{Res}I_{\mathfrak{C}^{\pm }}$.\newline

In any admissible basis, the entries of the matrices which represent the
relations (\ref{75}) and (\ref{76}) are expressed in much the same way as
those of (\ref{22}) and (\ref{23}). We have the defining constraints%
\begin{equation}
u_{\mu \lambda }^{++}g_{+}^{\ast \lambda \sigma }u_{\sigma \nu }^{++\bigstar
}+u_{\mu \lambda }^{+-}g_{-}^{\ast \lambda \sigma }u_{\sigma \nu
}^{+-\bigstar }=g_{\mu \nu }^{+}  \label{77}
\end{equation}%
and%
\begin{equation}
u_{\mu \lambda }^{--}g_{-}^{\ast \lambda \sigma }u_{\sigma \nu }^{--\bigstar
}+u_{\mu \lambda }^{-+}g_{+}^{\ast \lambda \sigma }u_{\sigma \nu
}^{-+\bigstar }=g_{\mu \nu }^{-},  \label{78}
\end{equation}%
along with%
\begin{equation}
u_{\mu \lambda }^{++}g_{+}^{\ast \lambda \sigma }u_{\sigma \nu }^{-+\bigstar
}+u_{\mu \lambda }^{+-}g_{-}^{\ast \lambda \sigma }u_{\sigma \nu
}^{--\bigstar }=0_{2}  \label{79}
\end{equation}%
and the $\bigstar $-conjugate of (\ref{79}). Obviously, the entries for $%
u^{+-\bigstar }$ and $u^{-+\bigstar }$ must account for the interchange of
operator actions borne by (\ref{8}) and (\ref{21}) whence, in the
computational basis, we have%
\begin{equation}
u_{\mu \nu }^{+-}=<e_{(\mu )}^{+}\mid u^{+-}\parallel e_{(\nu
)}^{-}>_{g^{-}},\text{ }u_{\mu \nu }^{+-\bigstar }=<e_{(\mu )}^{-}\mid
u^{+-\bigstar }\parallel e_{(\nu )}^{+}>_{g^{+}}  \label{81}
\end{equation}%
and%
\begin{equation}
u_{\mu \nu }^{-+}=<e_{(\mu )}^{-}\mid u^{-+}\parallel e_{(\nu
)}^{+}>_{g^{+}},\text{ }u_{\mu \nu }^{-+\bigstar }=<e_{(\mu )}^{+}\mid
u^{-+\bigstar }\parallel e_{(\nu )}^{-}>_{g^{-}}.  \label{82}
\end{equation}

Whenever $u$ is taken to bear unitarity as well, its decomposition as
provided by (\ref{75}) and (\ref{76}) turns out to be such that the
constituents $u^{+-}$and $u^{-+}$ amount both to zero operators. Under this
circumstance, we should take into account the conditions%
\begin{equation}
u^{++\bigstar }=u^{++\dagger },\text{ }u^{--\bigstar }=u^{--\dagger },
\label{83}
\end{equation}%
with%
\begin{equation}
\left( u_{\mu \nu }^{++}\right) \in GL(2,\mathbb{C})\cap U(2)\ni \left(
u_{\mu \nu }^{--}\right) ,\text{ }\func{det}\left( u_{\mu \nu }^{++}\right)
=\exp [i\phi ]=\func{det}\left( u_{\mu \nu }^{--}\right) ^{-1},  \label{84}
\end{equation}%
and $\phi $ being some real number. Therefore, any unitary element of $%
SU(2,2)$ possesses the block-diagonal form%
\begin{equation}
\left( u_{\mu \nu }\right) =\left( 
\begin{array}{ll}
\left( u_{\lambda \sigma }^{++}\right) & 0_{2} \\ 
0_{2} & \left( u_{\lambda \sigma }^{--}\right)%
\end{array}%
\right) .  \label{85}
\end{equation}%
From (\ref{22}) and (\ref{74}), we can see that the matrix $\left( I_{\mu
\nu }\right) $ bears $SU(2,2)$-invariance by definition, while $\left(
\Delta _{\mu \nu }\right) $ and $\left( \Delta ^{\ast \mu \nu }\right) $
must bear invariance under transformations belonging to the unitary
intersection $SU(2,2)\cap U(4).$

\section*{Traces}

The trace of an invertible operator $A$ in $\mathfrak{C}$ is a linear
attribute defined as%
\begin{equation}
\func{Tr}\text{ }A=<e_{(\mu )}\mid A\parallel e_{(\lambda )}>_{g}g^{\ast
\lambda \mu }=A_{\mu }{}^{\mu },  \label{86}
\end{equation}%
with its adjoint counterpart being%
\begin{equation}
\func{Tr}\text{ }A^{\ast }=<e^{\ast (\mu )}\mid A^{\ast }\parallel e^{\ast
(\lambda )}>_{g^{\ast }}g_{\lambda \mu }=A^{\ast \mu }{}_{\mu }.  \label{87}
\end{equation}%
More explicitly, we have%
\begin{equation}
\func{Tr}\text{ }A=A_{00}+A_{11}-A_{22}-A_{33},  \label{90}
\end{equation}%
or, equivalently,%
\begin{equation}
\func{Tr}\text{ }A=a_{00}+a_{11}+a_{22}+a_{33}.  \label{91}
\end{equation}%
Equations (\ref{linn}), (\ref{add1}) and (\ref{55}) thus supply us with the
restricted expressions%
\begin{equation}
\func{Tr}\text{ }A^{\pm }=A_{\mu \lambda }^{\pm }g_{\pm }^{\ast \lambda \mu
}=g_{\mu \lambda }^{\pm }A_{\pm }^{\ast \lambda \mu },\text{ }\func{Tr}\text{
}A_{\pm }^{\ast }=A_{\pm }^{\ast \mu \lambda }g_{\lambda \mu }^{\pm }=g_{\pm
}^{\ast \mu \lambda }A_{\lambda \mu }^{\pm },  \label{88}
\end{equation}%
which, by virtue of (\ref{49}), can be reset as%
\begin{equation}
\func{Tr}\text{ }A^{\pm }=a_{\mu \lambda }^{\pm }\Delta _{\pm }^{\ast
\lambda \mu },\text{ }\func{Tr}\text{ }A_{\pm }^{\ast }=a_{\pm }^{\ast \mu
\lambda }\Delta _{\lambda \mu }^{\pm }.  \label{89}
\end{equation}%
Then, for the decompositions (\ref{54}), we get%
\begin{equation}
\text{Tr }\mid e_{(\mu )}^{\pm }>A_{\pm }^{\ast \mu \nu }<e_{(\nu )}^{\pm
}\mid =\text{Tr }A^{\pm }  \label{tr1}
\end{equation}%
and%
\begin{equation}
\text{Tr }\mid e_{\pm }^{\ast (\mu )}>A_{\mu \nu }^{\pm }<e_{\pm }^{\ast
(\nu )}\mid =\text{Tr }A_{\pm }^{\ast }.  \label{tr2}
\end{equation}

By recalling (\ref{31}) and allowing for an operator product $AB$ along with
the entries%
\begin{equation}
<e_{(\mu )}^{\pm }\mid A^{\pm }B^{\pm }\parallel e_{(\nu )}^{\pm }>_{g^{\pm
}}=A_{\mu \lambda }^{\pm }g_{\pm }^{\ast \lambda \sigma }B_{\sigma \nu
}^{\pm }  \label{93}
\end{equation}%
and%
\begin{equation}
<e_{\pm }^{\ast (\mu )}\mid A_{\pm }^{\ast }B_{\pm }^{\ast }\parallel e_{\pm
}^{\ast (\nu )}>_{g_{\pm }^{\ast }}=A_{\pm }^{\ast \mu \lambda }g_{\lambda
\sigma }^{\pm }B_{\pm }^{\ast \sigma \nu },  \label{94}
\end{equation}%
we obtain%
\begin{equation}
\func{Tr}\text{ }(A^{\pm }B^{\pm })=A_{\mu \lambda }^{\pm }g_{\pm }^{\ast
\lambda \sigma }B_{\sigma \rho }^{\pm }g_{\pm }^{\ast \rho \mu }=A_{\mu
\lambda }^{\pm }B_{\pm }^{\ast \lambda \mu }  \label{95}
\end{equation}%
and%
\begin{equation}
\func{Tr}\text{ }\left( A_{\pm }^{\ast }B_{\pm }^{\ast }\right) =A_{\pm
}^{\ast \mu \lambda }g_{\lambda \sigma }^{\pm }B_{\pm }^{\ast \sigma \rho
}g_{\rho \mu }^{\pm }=A_{\pm }^{\ast \mu \lambda }B_{\lambda \mu }^{\pm }.
\label{96}
\end{equation}%
One could expect that the definition of traces of operators subjected to the
pseudo-Hermitian conjugation should be formally similar to (\ref{86}). This
is really the case as can be seen by invoking Eqs. (\ref{62}) and (\ref{63})
along with (\ref{64}) to perform the computation%
\begin{equation}
\text{Tr }A^{\pm \bigstar }=g_{\mu }^{\pm }{}^{\lambda }a_{\lambda \sigma
}^{\pm \dagger }g_{\pm }^{\ast \sigma \mu }=g_{\mu }^{\pm }{}^{\lambda
}A_{\lambda }^{\pm \dagger }{}^{\sigma }\Delta _{\sigma \rho }^{\pm }g_{\pm
}^{\ast \rho \mu }=g_{\mu }^{\pm }{}^{\lambda }A_{\lambda }^{\pm \dagger
}{}^{\sigma }g_{\sigma }^{\pm }{}^{\mu }=A_{\mu }^{\pm \bigstar }{}^{\mu }.
\label{92}
\end{equation}

The formal properties of traces of operators in Cartan's space are the same
as the ones which occur in the Hilbert-space framework [5,22]. A typical
unrestricted calculation is, in effect, carried out as follows%
\begin{equation}
<e_{(\mu )}\mid A\parallel e_{(\lambda )}>_{g}g^{\ast \lambda \mu }=%
\overline{<e_{(\lambda )}\mid A^{\bigstar }\parallel e_{(\mu )}>_{g}g^{\ast
\mu \lambda }}=\overline{A_{\lambda \mu }^{\bigstar }g^{\ast \mu \lambda }},
\label{97}
\end{equation}%
which brings about the relation%
\begin{equation}
\func{Tr}\text{ }A^{\pm }=\overline{\func{Tr}\text{ }A^{\pm \bigstar }}.
\label{98}
\end{equation}%
In a similar way, we obtain%
\begin{equation}
\func{Tr}\text{ }A_{\pm }^{\ast }=\overline{\func{Tr}\text{ }A_{\pm }^{\ast
\bigstar }},  \label{99}
\end{equation}%
whence, using (\ref{98}), say, with $AB$ in place of $A,$ yields%
\begin{equation}
\func{Tr}\text{ }(A^{\pm }B^{\pm })=\overline{\func{Tr}\text{ }\left( B^{\pm
\bigstar }A^{\pm \bigstar }\right) }.  \label{100}
\end{equation}%
We notice that Eq. (\ref{100}) can be straightforwardly obtained from the
combination of (\ref{98}) with the prescription%
\begin{equation}
\func{Tr}\text{ }(A^{\pm }B^{\pm })=\func{Tr}\text{ }(A^{\pm }B^{\pm
})^{\bigstar \bigstar }=\func{Tr}\text{ }\left( B^{\pm \bigstar }A^{\pm
\bigstar }\right) ^{\bigstar },  \label{101}
\end{equation}%
which enables us to carry out the development\footnote{%
It is worth remarking that the $\ast $-adjointness commutes with the $%
\bigstar $-conjugation.}%
\begin{align}
\func{Tr}\text{ }(A^{\pm }B^{\pm })& =\overline{\func{Tr}\text{ }(A^{\pm
}B^{\pm })^{\bigstar }}=\overline{\func{Tr}\text{ }\left( B^{\pm \bigstar
}A^{\pm \bigstar }\right) }=\overline{B_{\mu \lambda }^{\pm \bigstar }A_{\pm
}^{\ast \bigstar \lambda \mu }}  \notag \\
& =B_{\mu \lambda }^{\pm T}A_{\pm }^{\ast T\lambda \mu }=B_{\lambda \mu
}^{\pm }A_{\pm }^{\ast \mu \lambda }=\func{Tr}\text{ }(B^{\pm }A^{\pm }),
\label{102}
\end{align}%
whose adjoint version reads off%
\begin{equation}
\func{Tr}\text{ }\left( A_{\pm }^{\ast }B_{\pm }^{\ast }\right) =\func{Tr}%
\text{ }\left( B_{\pm }^{\ast }A_{\pm }^{\ast }\right) .  \label{103}
\end{equation}

If $B$ is reset as $A^{\bigstar }$ in (\ref{100}), we will conclude that%
\begin{equation}
\func{Im}\func{Tr}\text{ }\left( A^{\pm }A^{\pm \bigstar }\right) =0.
\label{104}
\end{equation}%
In the case of a pseudo-unitary $A$, we thus obtain the particular value%
\begin{equation}
\func{Tr}\text{ }\left( A^{\pm }A^{\pm \bigstar }\right) =\func{Tr}\text{ }%
\left( A^{\pm \bigstar }A^{\pm }\right) =\func{Tr}\text{ }I^{\pm }=g_{\mu
\lambda }^{\pm }g_{\pm }^{\ast \lambda \mu }=2,  \label{105}
\end{equation}%
whilst in the pseudo-Hermitian case, by (\ref{98}) and (\ref{104}), we get%
\begin{equation}
\func{Im}\func{Tr}\text{ }A^{\pm }=0,\text{ }\func{Im}\func{Tr}\text{ }%
(A^{\pm })^{2}=0.  \label{106}
\end{equation}%
Furthermore, if the operator $A$ carried by (\ref{105}) also holds
unitarity, then we may write%
\begin{equation}
\func{Tr}\text{ }A^{\pm \bigstar }=\func{Tr}\text{ }A^{\pm \dagger },
\label{107}
\end{equation}%
which produces the specialized configuration%
\begin{equation}
A_{\mu \lambda }^{\pm \bigstar }g_{\pm }^{\ast \lambda \mu }=a_{\mu \lambda
}^{\pm \dagger }\Delta _{\pm }^{\ast \lambda \mu }.  \label{108}
\end{equation}

\section*{Measurement processes}

We shall in this Section transcribe into the framework of Cartan's space the
description of the measurement processes we had referred to before. It will
suffice to allow for the unstarred picture of Section 2 as far as our
transcription is concerned, since adjoint formulations in Cartan's space are
physically equivalent.

According to the usual definition of complete sets of commuting observables
[1,2,5], any quantum system may have in some given frame only one complete
set of compatible observables and these can always admit common eigenvectors
even within the context of indefinite inner-product spaces, as mentioned in
Section 1. We consider an ensemble constituted by pairs of companion systems%
\begin{equation}
\{(\mathcal{A}^{+},\mathcal{A}^{-}),(\mathcal{B}^{+},\mathcal{B}^{-}),(%
\mathcal{C}^{+},\mathcal{C}^{-}),...,(\mathcal{D}^{+},\mathcal{D}^{-})\},
\label{m5}
\end{equation}%
with the complete sets for each pair $(\Sigma ^{+},\Sigma ^{-})$ carrying
observables that supposedly possess non-degenerate twofold discrete spectra
and can be measured separately. The complete sets for different pairs should
carry observables that act on identifiable copies of Cartan's space such as,
for instance, $\mathfrak{C}^{+}(\mathcal{A}^{+})$ and $\mathfrak{C}^{-}(%
\mathcal{B}^{-})$, whereas the space of states for the ensemble should be
looked upon as the tensor product of all copies, namely\footnote{%
Tensor products of copies of Cartan's space involve juxtapositions of
suitable numbers of $g^{\pm }$ and $g_{\pm }^{\ast }.$ A detailed
description of composite systems in Cartan's space can be found in Section 4
of Ref. [13].}%
\begin{equation}
\mathfrak{C}^{+}(\mathcal{A}^{+})\otimes \mathfrak{C}^{-}(\mathcal{A}%
^{-})\otimes ...\otimes \mathfrak{C}^{+}(\mathcal{D}^{+})\otimes \mathfrak{C}%
^{-}(\mathcal{D}^{-}).  \label{prod1}
\end{equation}

The common eigenbras of the observables for $\Sigma ^{\pm }$ are taken as
the elements of the computational basis of the copy $\mathfrak{C}^{\pm
}(\Sigma ^{\pm })$ but, as the computational bases of all copies are
indistinguishable among themselves, we will for simplicity denote the common
eigenbras for the complete sets of all systems of (\ref{m5}) as\footnote{%
By this point, we recall that (\ref{34}) entails that the entry labels borne
by any restricted configurations must just take the values 0 and 1. The
notation of (\ref{111}) may formally facilitate the evaluation of inner
products between states for different systems of (\ref{m5}).}%
\begin{equation}
\{<e_{(0)}^{+}\mid ,<e_{(1)}^{+}\mid \},\text{ }\{<e_{(0)}^{-}\mid
,<e_{(1)}^{-}\mid \}.  \label{111}
\end{equation}%
A state bra for $\Sigma ^{\pm }$ is identified with a normalized element of $%
\mathfrak{C}^{\pm }(\Sigma ^{\pm }),$ and thus written as%
\begin{equation}
<s^{\pm }\mid =S_{\pm }^{\mu }<e_{(\mu )}^{\pm }\mid ,  \label{115}
\end{equation}%
with Eqs. (\ref{45}) and (\ref{46}) yielding the amplitudes%
\begin{equation}
S_{\pm }^{0}=\pm <s^{\pm }\mid e_{(0)}^{\pm }>_{g^{\pm }},\text{ }S_{\pm
}^{1}=\pm <s^{\pm }\mid e_{(1)}^{\pm }>_{g^{\pm }}  \label{add52}
\end{equation}%
and the normalization condition%
\begin{equation}
<s^{\pm }\mid g^{\pm }\parallel s^{\pm }>_{g^{\pm }}=\text{Tr}\mid s^{\pm
}><s^{\pm }\mid g^{\pm }=S_{\pm }^{\mu }\Delta _{\mu \nu }^{\pm }\overline{%
S_{\pm }^{\nu }}=1,  \label{add51}
\end{equation}%
such that we can write the density equations%
\begin{equation}
<s^{\pm }\mid \rho (\Sigma ^{\pm })=<s^{\pm }\mid ,\text{ }\rho ^{\bigstar
}(\Sigma ^{\pm })\mid s^{\pm }>=\mid s^{\pm }>,\text{ }  \label{j1}
\end{equation}%
where\footnote{%
According to Ref. [13], the local evolution of any state is controlled by a
suitably defined unitary operator restriction which has to be required to
depend explicitly only upon the proper time.}%
\begin{equation}
\rho (\Sigma ^{\pm })=g^{\pm }\mid s^{\pm }><s^{\pm }\mid =\mid s^{\pm
}><s^{\pm }\mid g^{\pm }=\rho ^{\bigstar }(\Sigma ^{\pm }).  \label{j2}
\end{equation}

We adopt the definition put into practice in Ref. [13] by which any
observables belonging to the complete set for either $\Sigma ^{+}$ or $%
\Sigma ^{-}$ amount to pseudo-Hermitian restrictions on $\mathfrak{C}^{\pm
}(\Sigma ^{\pm })$ that formally resemble the operator pieces carried by Eq.
(\ref{linn}). Then, by (\ref{59}), such an observable $S^{\pm }$ must obey
the Hermiticity relations%
\begin{equation}
S^{\pm }=S^{\pm \bigstar }=S^{\pm \dagger },  \label{add15}
\end{equation}%
which guarantee [19,21] its spectral reality, whereas%
\begin{equation}
D(S^{\pm })=R(S^{\pm })=\mathfrak{C}^{\pm }(\Sigma ^{\pm }).  \label{add 21}
\end{equation}%
The spectra 
\begin{equation}
\text{spec }S^{+}=\left\{ s_{0}^{+},s_{1}^{+}\},\text{ spec }%
S^{-}=\{s_{0}^{-},s_{1}^{-}\right\} ,  \label{spec1}
\end{equation}%
come straightaway from the eigenvalue equations%
\begin{equation}
<e_{(0)}^{\pm }\mid S^{\pm }=s_{0}^{\pm }<e_{(0)}^{\pm }\mid ,\text{ }%
<e_{(1)}^{\pm }\mid S^{\pm }=s_{1}^{\pm }<e_{(1)}^{\pm }\mid ,  \label{m1}
\end{equation}%
which produce the spectral entries%
\begin{equation}
<e_{(0)}^{\pm }\mid S^{\pm }\parallel e_{(0)}^{\pm }>_{g^{\pm }}=\pm
s_{0}^{\pm }  \label{m2}
\end{equation}%
and%
\begin{equation}
<e_{(1)}^{\pm }\mid S^{\pm }\parallel e_{(1)}^{\pm }>_{g^{\pm }}=\pm
s_{1}^{\pm },  \label{m2lin}
\end{equation}%
together with the diagonal array%
\begin{equation}
(S_{\mu \nu }^{\pm })=\left( 
\begin{array}{ll}
\pm s_{0}^{\pm } & 0 \\ 
0 & \pm s_{1}^{\pm }%
\end{array}%
\right)  \label{700}
\end{equation}%
and the expansion%
\begin{equation}
<s^{\pm }\mid S^{\pm }=S_{\pm }^{0}s_{0}^{\pm }<e_{(0)}^{\pm }\mid +\text{ }%
S_{\pm }^{1}s_{1}^{\pm }<e_{(1)}^{\pm }\mid .  \label{exp}
\end{equation}%
Consequently, the spectral decompositions (\ref{54}) yield the definite
traces%
\begin{equation}
\func{Tr}\text{ }S^{\pm }=\func{Tr}\mid e_{(\mu )}^{\pm }>S_{\pm }^{\ast \mu
\nu }<e_{(\nu )}^{\pm }\mid =s_{0}^{\pm }+s_{1}^{\pm },  \label{109}
\end{equation}%
whence, for the pertinent expectation value of $S^{\pm },$ we have%
\begin{equation}
<S^{\pm }>_{s^{\pm }}=\pm s_{\mu }^{\pm }\mid S_{\pm }^{\mu }\mid ^{2}=\text{
Tr }(\rho (\Sigma ^{\pm })S^{\pm }).  \label{300}
\end{equation}

The full apparatuses used for performing measurements on quantum systems
generally consist of ordered sequences of physical devices that pick up and
transmit state vectors out of incoming ones, in agreement with prescribed
dynamical selections. In the works exhibited in Refs. [2,3], each such
device is represented by a Hermitian operator in the Hilbert space of a
non-relativistic microcospic ensemble whose dynamical attributes possess
finite numbers of eigenstates. Incoming states of similar systems are thus
sorted by the utilized measurement devices into outgoing subensembles that
are distinguishable from each other by means of selected eigenstates.

We require any measurement device to be represented by a decomposable
operator restriction on an appropriate subspace of the tensor product (\ref%
{prod1}). Selective measurements on the systems (\ref{m5}) are accomplished
from the implementation of measurement operators that adequately absorb
amplitudes of state vectors upon performing state selections and reductions.

The simplest measurement process involves a selective measurement on some $%
\Sigma ^{\pm }$ which is in either case performed by a single device that
accepts one of the eigenbras carried by (\ref{111}) while correspondingly
rejecting the other. For any such process, the representative measurement
operator appears as one of the restrictions%
\begin{equation}
\pi _{(0)}^{+}(\Sigma ^{+}),\text{ }\pi _{(1)}^{+}(\Sigma ^{+}),\text{ }\pi
_{(0)}^{-}(\Sigma ^{-}),\text{ }\pi _{(1)}^{-}(\Sigma ^{-}),  \label{m12}
\end{equation}%
which, by definition, bears pseudo-Hermiticity and reduces an incoming state
in accordance with the prescription%
\begin{equation}
<s^{\pm }\mid \pi _{(\mu )}^{\pm }(\Sigma ^{\pm })=<s_{(\mu )}^{\pm }\mid ,
\label{m9}
\end{equation}%
where\footnote{%
The eigenvalue character of (\ref{m1}) will still hold if we substitute $%
<e_{(\mu )}^{\pm }\mid $ for $<s_{(\mu )}^{\pm }\mid .$}%
\begin{equation}
<s_{(0)}^{\pm }\mid =S_{\pm }^{0}<e_{(0)}^{\pm }\mid ,\text{ }<s_{(1)}^{\pm
}\mid =S_{\pm }^{1}<e_{(1)}^{\pm }\mid .  \label{m10}
\end{equation}%
We can then write%
\begin{equation}
<s^{\pm }\mid =<s_{(0)}^{\pm }\mid +<s_{(1)}^{\pm }\mid  \label{state1}
\end{equation}%
and%
\begin{equation}
<s^{\pm }\mid s^{\pm }>_{g^{\pm }}=<s_{(0)}^{\pm }\mid s_{(0)}^{\pm
}>_{g^{\pm }}+<s_{(1)}^{\pm }\mid s_{(1)}^{\pm }>_{g^{\pm }},  \label{m50}
\end{equation}%
such that $\pi _{(\mu )}^{\pm }(\Sigma ^{\pm })$ also causes the
reduced-state annihilations%
\begin{equation}
<s_{(1)}^{\pm }\mid \pi _{(0)}^{\pm }(\Sigma ^{\pm })=<0^{\pm }\mid ,\text{ }%
<s_{(0)}^{\pm }\mid \pi _{(1)}^{\pm }(\Sigma ^{\pm })=<0^{\pm }\mid ,
\label{m51}
\end{equation}%
while (\ref{exp}) becomes%
\begin{equation}
<s^{\pm }\mid S^{\pm }=s_{0}^{\pm }<s_{(0)}^{\pm }\mid +\text{ }s_{1}^{\pm
}<s_{(1)}^{\pm }\mid .  \label{exp1}
\end{equation}%
Equations (\ref{m9}) and (\ref{m10}) thus give the completeness decomposition%
\begin{equation}
\pi _{(\mu )}^{\pm }(\Sigma ^{\pm })=\mid e_{(\rho )}^{\pm }>g_{\pm }^{\ast
\rho \lambda }\pi _{(\mu )\lambda \sigma }^{\pm }(\Sigma ^{\pm })g_{\pm
}^{\ast \sigma \tau }<e_{(\tau )}^{\pm }\mid ,  \label{J90}
\end{equation}%
along with the representations\footnote{%
Matrix representations equal their adjoint counterparts in accordance with
Eqs. (\ref{52})-(\ref{55}).}%
\begin{equation}
(\pi _{(0)\lambda \sigma }^{\pm }{}(\Sigma ^{\pm }))=%
\begin{pmatrix}
\pm 1 & 0 \\ 
0 & 0%
\end{pmatrix}%
,\text{ }(\pi _{(1)\lambda \sigma }^{\pm }{}(\Sigma ^{\pm }))=%
\begin{pmatrix}
0 & 0 \\ 
0 & \pm 1%
\end{pmatrix}
\label{m15}
\end{equation}%
and%
\begin{equation}
\pi _{(\mu )\lambda \sigma }^{\pm }{}(\Sigma ^{\pm })=<e_{(\lambda )}^{\pm
}\mid \pi _{(\mu )}^{\pm }(\Sigma ^{\pm })\parallel e_{(\sigma )}^{\pm
}>_{g^{\pm }},  \label{m16}
\end{equation}%
whence, fitting together the first of (\ref{88}) with (\ref{m15}), yields%
\begin{equation}
\text{Tr }\pi _{(\mu )}^{\pm }{}(\Sigma ^{\pm })=1,  \label{T3}
\end{equation}%
whilst the operators (\ref{m12}) hold the projection property\footnote{%
The operators $\pi _{(\mu )}^{\pm }$ may afford a relativistic version of
POVM operators.}%
\begin{equation}
\pi _{(0)}^{\pm }(\Sigma ^{\pm })\pi _{(0)}^{\pm }(\Sigma ^{\pm })=\pi
_{(0)}^{\pm }(\Sigma ^{\pm }),\text{ }\pi _{(1)}^{\pm }(\Sigma ^{\pm })\pi
_{(1)}^{\pm }(\Sigma ^{\pm })=\pi _{(1)}^{\pm }(\Sigma ^{\pm }).
\label{add90}
\end{equation}%
The state reductions (\ref{m9}) and the occurrences of the spectral values
as set by (\ref{spec1}), are interrelated via the Born probabilities%
\begin{equation}
<s^{\pm }\mid \pi _{(0)}^{\pm }(\Sigma ^{\pm })g^{\pm }\parallel s^{\pm
}>_{g^{\pm }}=\mid S_{\pm }^{0}\mid ^{2}  \label{m}
\end{equation}%
and 
\begin{equation}
<s^{\pm }\mid \pi _{(1)}^{\pm }(\Sigma ^{\pm })g^{\pm }\parallel s^{\pm
}>_{g^{\pm }}=\mid S_{\pm }^{1}\mid ^{2},  \label{mlin}
\end{equation}%
with one of Eqs. (\ref{62}) having been employed to write (\ref{m}) and (\ref%
{mlin}). Hence, the condition (\ref{add51}) can be reinstated as%
\begin{equation}
<s^{\pm }\mid g^{\pm }\parallel s^{\pm }>_{g^{\pm }}=<<s^{\pm }\mid \pi
_{(0)}^{\pm }(\Sigma ^{\pm })\parallel s^{\pm }>>+<<s^{\pm }\mid \pi
_{(1)}^{\pm }(\Sigma ^{\pm })\parallel s^{\pm }>>,  \label{m91}
\end{equation}%
or, alternatively, as%
\begin{equation}
\text{Tr}\mid e_{(\lambda )}^{\pm }>\pi _{\pm }^{\ast (\mu )\lambda \sigma
}{}(\Sigma ^{\pm })<e_{(\sigma )}^{\pm }\mid =1,  \label{m100}
\end{equation}%
but the normalizability of $<s^{\pm }\mid $ gets lost when the reductions
prescribed by $\pi _{(\mu )}^{\pm }{}(\Sigma ^{\pm })$ are brought into
effect. Therefore, the significant reduced states produced by the
measurement operators (\ref{m12}), are given by%
\begin{equation}
\frac{1}{\mid S_{\pm }^{0}\mid }<s_{(0)}^{\pm }\mid ,\text{ }\frac{1}{\mid
S_{\pm }^{1}\mid }<s_{(1)}^{\pm }\mid .  \label{m101}
\end{equation}

The process just described can be extended to the\ extent that a measurement
operator $\pi ^{\pm }(s_{\mu }^{\pm },s_{\nu }^{\pm })$ is used\footnote{%
Our non-degeneracy assumption allows us to unambiguously identify
measurement selections and reductions by means of eigenvalues.} in lieu of
those of (\ref{m12}), which annihilates an incoming reduced state $<s_{(\mu
)}^{\pm }\mid $ and produces an outgoing reduced state $<s_{(\nu )}^{\pm
}\mid .$ It is clear that the situation of general interest demands the
label distinction $\mu \neq \nu ,$ such that the operator $\pi ^{\pm
}(s_{\mu }^{\pm },s_{\nu }^{\pm })$ generally should perform the
simultaneous interchanges%
\begin{equation}
0^{\pm }\longleftrightarrow 1^{\pm }.  \label{m103}
\end{equation}%
Of course, as regards this latter process, some incoming state could have
been prepared beforehand by a device from (\ref{m12}), whereas the
respective outgoing state would emerge as another prepared state from the
operationality of $\pi ^{\pm }(s_{\mu }^{\pm },s_{\nu }^{\pm }).$ We have
the operator-eigenvalue association%
\begin{equation}
\pi ^{\pm }(s_{\mu }^{\pm },s_{\nu }^{\pm })\longmapsto \pi ^{\pm }(\pi
_{(\mu )}^{\pm }(\Sigma ^{\pm }),\pi _{(\nu )}^{\pm }(\Sigma ^{\pm })),
\label{m102}
\end{equation}%
along with the annihilation-production prescriptions%
\begin{equation}
<s_{(0)}^{\pm }\mid \pi ^{\pm }(s_{1}^{\pm },s_{0}^{\pm })=<0^{\pm }\mid ,%
\text{ }<s_{(1)}^{\pm }\mid \pi ^{\pm }(s_{0}^{\pm },s_{1}^{\pm })=<0^{\pm
}\mid  \label{Add10}
\end{equation}%
and%
\begin{equation}
<s_{(0)}^{\pm }\mid \pi ^{\pm }(s_{0}^{\pm },s_{1}^{\pm })=<s_{(1)}^{\pm
}\mid ,\text{ }<s_{(1)}^{\pm }\mid \pi ^{\pm }(s_{1}^{\pm },s_{0}^{\pm
})=<s_{(0)}^{\pm }\mid .  \label{Add9}
\end{equation}%
Hence, as Eq. (\ref{m10}) ensures that any reduced states carrying different
labels must bear $g^{\pm }$-orthogonality, we may write the non-vanishing
products%
\begin{equation}
<s_{(0)}^{\pm }\mid \pi ^{\pm }(s_{0}^{\pm },s_{1}^{\pm })\parallel
s_{(1)}^{\pm }>_{g^{\pm }}=<s_{(1)}^{\pm }\mid s_{(1)}^{\pm }>_{g^{\pm
}}=\pm \mid S_{\pm }^{1}\mid ^{2}  \label{1*}
\end{equation}%
and%
\begin{equation}
<s_{(1)}^{\pm }\mid \pi ^{\pm }(s_{1}^{\pm },s_{0}^{\pm })\parallel
s_{(0)}^{\pm }>_{g^{\pm }}=<s_{(0)}^{\pm }\mid s_{(0)}^{\pm }>_{g^{\pm
}}=\pm \mid S_{\pm }^{0}\mid ^{2},  \label{2*}
\end{equation}%
which agree with both (\ref{m50}) and%
\begin{equation}
<s^{\pm }\mid s^{\pm }>_{g^{\pm }}=<s^{\pm }\mid \pi ^{\pm }(s_{0}^{\pm
},s_{1}^{\pm })\parallel s^{\pm }>_{g^{\pm }}+<s^{\pm }\mid \pi ^{\pm
}(s_{1}^{\pm },s_{0}^{\pm })\parallel s^{\pm }>_{g^{\pm }}.  \label{*}
\end{equation}%
Equations (\ref{Add10})-(\ref{2*}) thus give the decomposition%
\begin{equation}
\pi ^{\pm }(s_{\mu }^{\pm },s_{\nu }^{\pm })=\mid e_{(\lambda )}^{\pm }>\pi
_{\pm }^{\ast \lambda \sigma }(s_{\mu }^{\pm },s_{\nu }^{\pm })<e_{(\sigma
)}^{\pm }\mid ,  \label{9*}
\end{equation}%
along with the entries%
\begin{equation}
<e_{(0)}^{\pm }\mid \pi ^{\pm }(s_{0}^{\pm },s_{1}^{\pm })\parallel
e_{(1)}^{\pm }>_{g^{\pm }}=\pm \frac{S^{1}}{S^{0}}  \label{5*}
\end{equation}%
and%
\begin{equation}
<e_{(1)}^{\pm }\mid \pi ^{\pm }(s_{1}^{\pm },s_{0}^{\pm })\parallel
e_{(0)}^{\pm }>_{g^{\pm }}=\pm \frac{S^{0}}{S^{1}},  \label{5**}
\end{equation}%
which constitute the representations%
\begin{equation}
(\pi _{\mu \nu }^{\pm }(s_{0}^{\pm },s_{1}^{\pm }))=\left( 
\begin{array}{cc}
0 & \pm \frac{S^{1}}{S^{0}} \\ 
0 & 0%
\end{array}%
\right) ,\text{ }(\pi _{\mu \nu }^{\pm }(s_{1}^{\pm },s_{0}^{\pm }))=\left( 
\begin{array}{cc}
0 & 0 \\ 
\pm \frac{S^{0}}{S^{1}} & 0%
\end{array}%
\right) .  \label{3*}
\end{equation}%
The $\bigstar $-conjugates of (\ref{Add10}) and (\ref{Add9}) should then
absorb an interchange of reduced-state labels, that obeys the associated
arrangements%
\begin{equation}
\pi ^{\pm \bigstar }(s_{0}^{\pm },s_{1}^{\pm })\mid s_{(0)}^{\pm }>=\mid
0^{\pm }>,\text{ }\pi ^{\pm \bigstar }(s_{1}^{\pm },s_{0}^{\pm })\mid
s_{(1)}^{\pm }>=\mid 0^{\pm }>  \label{a}
\end{equation}%
and%
\begin{equation}
\pi ^{\pm \bigstar }(s_{1}^{\pm },s_{0}^{\pm })\mid s_{(0)}^{\pm }>=\mid
s_{1}^{\pm }>,\text{ }\pi ^{\pm \bigstar }(s_{0}^{\pm },s_{1}^{\pm })\mid
s_{(1)}^{\pm }>=\mid s_{0}^{\pm }>,  \label{c}
\end{equation}%
that is to say%
\begin{equation}
<s_{(0)}^{\pm }\mid \pi ^{\pm }(s_{0}^{\pm },s_{1}^{\pm })\parallel
s_{(1)}^{\pm }>_{g^{\pm }}=<s_{(1)}^{\pm }\parallel \pi ^{\pm \bigstar
}(s_{1}^{\pm },s_{0}^{\pm })\mid s_{(0)}^{\pm }>_{g^{\pm }}  \label{e}
\end{equation}%
and%
\begin{equation}
<s_{(1)}^{\pm }\mid \pi ^{\pm }(s_{1}^{\pm },s_{0}^{\pm })\parallel
s_{(0)}^{\pm }>_{g^{\pm }}=<s_{(0)}^{\pm }\parallel \pi ^{\pm \bigstar
}(s_{0}^{\pm },s_{1}^{\pm })\mid s_{(1)}^{\pm }>_{g^{\pm }},  \label{g}
\end{equation}%
which surely conform to the $\bigstar $-version of (\ref{3*}).

Let us now allow for a process related to a sequence of pseudo-Hermitian
measurement operators\footnote{%
Any sequence of operators to be considered hereafter should conventionally
be taken from left to right. This convention is different from the one
adopted in Ref. [2]. Thus the attributes "incoming" and "outgoing" always
get interchanged when the ordering convention is modified.}%
\begin{equation}
\Pi _{(\mu )}^{\pm }(\mathcal{B}^{\pm })\Pi _{(\lambda )}^{\pm }(\mathcal{A}%
^{\pm })\Pi _{(\sigma )}^{\pm }(\mathcal{B}^{\pm }),  \label{m6}
\end{equation}%
with each of which being defined so as to enter the decomposition%
\begin{equation}
\Pi ^{\pm }(\Sigma ^{\pm })=\mid s_{(\alpha )}^{\pm }>g_{\pm }^{\ast \alpha
\beta }<s_{(\beta )}^{\pm }\mid =\Pi _{(0)}^{\pm }(\Sigma ^{\pm })+\Pi
_{(1)}^{\pm }(\Sigma ^{\pm }),  \label{add102}
\end{equation}%
where\footnote{%
The operators $\mid s_{(0)}^{\pm }><s_{(1)}^{\pm }\mid $ and $\mid
s_{(1)}^{\pm }><s_{(0)}^{\pm }\mid $ bear tracelessness and have been ruled
out here.}%
\begin{equation}
\Pi _{(0)}^{\pm }(\Sigma ^{\pm })=\pm \mid s_{(0)}^{\pm }><s_{(0)}^{\pm
}\mid =\pm \text{ }\pi _{(0)}^{\pm }(\Sigma ^{\pm })\mid s^{\pm }><s^{\pm
}\mid \pi _{(0)}^{\pm }(\Sigma ^{\pm })  \label{pi}
\end{equation}%
and%
\begin{equation}
\Pi _{(1)}^{\pm }(\Sigma ^{\pm })=\pm \mid s_{(1)}^{\pm }><s_{(1)}^{\pm
}\mid =\pm \text{ }\pi _{(1)}^{\pm }(\Sigma ^{\pm })\mid s^{\pm }><s^{\pm
}\mid \pi _{(1)}^{\pm }(\Sigma ^{\pm }),  \label{Pi}
\end{equation}%
with the kernel letters $\Sigma ,$ $s$ thus standing for either $\mathcal{B}%
, $ $b$ or $\mathcal{A},$ $a$. Presumably, the initial incoming state for
the process is prepared by $\Pi _{(\mu )}^{\pm }(\mathcal{B}^{\pm })$ as $%
<b_{(\mu )}^{\pm }a^{\pm }\mid $ from an impinging outer-product state like
either $<b^{+}a^{+}\mid $ or $<b^{-}a^{-}\mid .$ Also, (\ref{add102}) is
taken to involve the same reduced states as those occurring in (\ref{m9}),
whence we have the representations%
\begin{equation}
(\Pi _{(0)\gamma \delta }^{\pm }{}(\Sigma ^{\pm }))=%
\begin{pmatrix}
\pm \mid S_{\pm }^{0}\mid ^{2} & \text{ \ \ }0 \\ 
0 & \text{ \ \ }0%
\end{pmatrix}%
,\text{ }(\Pi _{(1)\gamma \delta }^{\pm }{}(\Sigma ^{\pm }))=%
\begin{pmatrix}
0 & \text{ \ \ }0 \\ 
0 & \text{ \ \ }\pm \mid S_{\pm }^{1}\mid ^{2}%
\end{pmatrix}%
,  \label{C}
\end{equation}%
along with the traces%
\begin{equation}
\text{Tr }\Pi _{(\eta )}^{\pm }{}(\Sigma ^{\pm })=\mid S_{\pm }^{\gamma
}g_{\gamma \eta }^{\pm }\mid ^{2},\text{ Tr }\Pi ^{\pm }{}(\Sigma ^{\pm })=%
\text{Tr }(\Pi _{(0)}^{\pm }{}(\Sigma ^{\pm })+\Pi _{(1)}^{\pm }{}(\Sigma
^{\pm }))=1.  \label{j}
\end{equation}%
Equation (\ref{m6}) can therefore be rewritten as%
\begin{equation}
\mid b_{(\mu )}^{\pm }>g_{\pm }^{\ast \mu \alpha }<b_{(\alpha )}^{\pm }\mid
a_{(\lambda )}^{\pm }>_{g^{\pm }}g_{\pm }^{\ast \lambda \beta }<a_{(\beta
)}^{\pm }\mid b_{(\sigma )}^{\pm }>_{g^{\pm }}g_{\pm }^{\ast \sigma \rho
}<b_{(\rho )}^{\pm }\mid ,  \label{J}
\end{equation}%
while (\ref{add102}) establishes that no crossed terms do really occur in (%
\ref{J}) for each value of the label carried by the prepared incoming state.
We then obtain the individual configurations from (\ref{J})%
\begin{equation}
\text{for }\mu =0\text{: }\mid <b_{(0)}^{\pm }\mid a_{(0)}^{\pm }>_{g^{\pm
}}\mid ^{2}\mid b_{(0)}^{\pm }>g_{\pm }^{\ast 00}<b_{(0)}^{\pm }\mid
\label{J1}
\end{equation}%
and%
\begin{equation}
\text{for }\mu =1\text{: }\mid <b_{(1)}^{\pm }\mid a_{(1)}^{\pm }>_{g^{\pm
}}\mid ^{2}\mid b_{(1)}^{\pm }>g_{\pm }^{\ast 11}<b_{(1)}^{\pm }\mid ,
\label{J2}
\end{equation}%
which suggest writing out the equality [2]%
\begin{equation}
\Pi _{(\mu )}^{\pm }(\mathcal{B}^{\pm })\Pi _{(\lambda )}^{\pm }(\mathcal{A}%
^{\pm })\Pi _{(\sigma )}^{\pm }(\mathcal{B}^{\pm })=\mid <b_{(\mu )}^{\pm
}\mid a_{(\lambda )}^{\pm }>_{g^{\pm }}\mid ^{2}\Pi _{(\sigma )}^{\pm }(%
\mathcal{B}^{\pm }),  \label{J3}
\end{equation}%
provided that $\mu =\lambda =\sigma .$ The operational effect of (\ref{m6})
differs from that of (\ref{add90}), with $\Sigma ^{\pm }$ being thereabout
identified with $\mathcal{B}^{\pm },$ because of the disturbance induced by
the $\mathcal{A}^{\pm }$-measurement [2] via the presence of $\Pi _{(\lambda
)}^{\pm }(\mathcal{A}^{\pm })$ in the second stage of the process, which
actually permits a fraction of $<b_{(\mu )}^{\pm }\mid $ to be passed on, in
addition to producing the $\mathcal{A}^{\pm }$-reduction $<a_{(\lambda
)}^{\pm }\mid $. Subsequently, the device $\Pi _{(\sigma )}^{\pm }(\mathcal{B%
}^{\pm })$ annihilates the reduced state $<b_{(\mu )}^{\pm }\mid $ unless $%
\mu =\sigma ,$ and allows for a fraction of $<a_{(\lambda )}^{\pm }\mid $ to
be transferred such that, when $\mu =\sigma $, the state $<b_{(\sigma
)}^{\pm }a_{(\lambda )}^{\pm }\mid $ will amount to what is transmitted
through the whole apparatus, with the relevant statistical transmission
weights being%
\begin{equation}
<b_{(0)}^{\pm }\mid a_{(0)}^{\pm }>_{g^{\pm }}<a_{(0)}^{\pm }\mid
b_{(0)}^{\pm }>_{g^{\pm }}=\mid <b_{(0)}^{\pm }\mid a_{(0)}^{\pm }>_{g^{\pm
}}\mid ^{2}  \label{J9}
\end{equation}%
and%
\begin{equation}
<b_{(1)}^{\pm }\mid a_{(1)}^{\pm }>_{g^{\pm }}<a_{(1)}^{\pm }\mid
b_{(1)}^{\pm }>_{g^{\pm }}=\mid <b_{(1)}^{\pm }\mid a_{(1)}^{\pm }>_{g^{\pm
}}\mid ^{2}.  \label{J10}
\end{equation}%
If we reexpress (\ref{m6}) by replacing $\Pi _{(\lambda )}^{\pm }(\mathcal{A}%
^{\pm })$ with $\pi _{(\lambda )}^{\pm }(\mathcal{A}^{\pm }),$ then the
corresponding modified version of (\ref{J}) would hide the $\mathcal{A}^{\pm
}$-states according to the Born-like description%
\begin{equation}
\Pi _{(\mu )}^{\pm }(\mathcal{B}^{\pm })\pi _{(\lambda )}^{\pm }(\mathcal{A}%
^{\pm })\Pi _{(\sigma )}^{\pm }(\mathcal{B}^{\pm })=<b_{(\mu )}^{\pm }\mid
b_{(\lambda )}^{\pm }>_{g^{\pm }}\Pi _{(\sigma )}^{\pm }(\mathcal{B}^{\pm }),
\label{J910}
\end{equation}%
with $\mu =\lambda =\sigma $ once again.

The sequence\footnote{%
The defining pseudo-Hermiticity of $\Pi _{(\mu )}^{\pm }(\Sigma ^{\pm })$
tells us that the arrangement (\ref{J15}) is invariant under the combination
of the $\bigstar $-conjugation with a kernel-letter interchange.}%
\begin{equation}
\Pi _{(\mu )}^{\pm }(\mathcal{A}^{\pm })\Pi _{(\lambda )}^{\pm }(\mathcal{B}%
^{\pm })\Pi _{(\sigma )}^{\pm }(\mathcal{C}^{\pm }),  \label{J15}
\end{equation}%
constitutes an extension of (\ref{m6}). Some manipulations similar to those
performed for the case of (\ref{m6}), give the arrangements%
\begin{equation}
\Pi _{(0)}^{\pm }(\mathcal{A}^{\pm })\Pi _{(0)}^{\pm }(\mathcal{B}^{\pm
})\Pi _{(0)}^{\pm }(\mathcal{C}^{\pm })=<a_{(0)}^{\pm }\mid b_{(0)}^{\pm
}>_{g^{\pm }}<b_{(0)}^{\pm }\mid c_{(0)}^{\pm }>_{g^{\pm }}M^{\pm
}(a_{0}^{\pm },c_{0}^{\pm })  \label{J91}
\end{equation}%
and%
\begin{equation}
\Pi _{(1)}^{\pm }(\mathcal{A}^{\pm })\Pi _{(1)}^{\pm }(\mathcal{B}^{\pm
})\Pi _{(1)}^{\pm }(\mathcal{C}^{\pm })=<a_{(1)}^{\pm }\mid b_{(1)}^{\pm
}>_{g^{\pm }}<b_{(1)}^{\pm }\mid c_{(1)}^{\pm }>_{g^{\pm }}M^{\pm
}(a_{1}^{\pm },c_{1}^{\pm }),  \label{J93}
\end{equation}%
along with the definitions%
\begin{equation}
M^{\pm }(a_{0}^{\pm },c_{0}^{\pm })=\pm \mid a_{(0)}^{\pm }><c_{(0)}^{\pm
}\mid ,\text{ }M^{\pm }(a_{1}^{\pm },c_{1}^{\pm })=\pm \mid a_{(1)}^{\pm
}><c_{(1)}^{\pm }\mid  \label{d5}
\end{equation}%
and the extended decomposition%
\begin{equation}
M^{\pm }(\mathcal{A}^{\pm },\mathcal{C}^{\pm })=\mid a_{(\alpha )}^{\pm
}>g_{\pm }^{\ast \alpha \beta }<c_{(\beta )}^{\pm }\mid =M^{\pm }(a_{0}^{\pm
},c_{0}^{\pm })+M^{\pm }(a_{1}^{\pm },c_{1}^{\pm }).  \label{M50}
\end{equation}%
We thus have the representations%
\begin{equation}
(M_{\mu \nu }^{\pm }(a_{0}^{\pm },c_{0}^{\pm }))=\left( 
\begin{array}{cc}
\pm C_{\pm }^{0}\overline{A_{\pm }^{0}} & 0 \\ 
0 & 0%
\end{array}%
\right) ,\text{ }(M_{\mu \nu }^{\pm }(a_{1}^{\pm },c_{1}^{\pm }))=\left( 
\begin{array}{cc}
0 & 0 \\ 
0 & \pm C_{\pm }^{1}\overline{A_{\pm }^{1}}%
\end{array}%
\right) ,  \label{M15}
\end{equation}%
which recover (\ref{C}) in the case of (\ref{m6}), while Eqs. (\ref{d5})-(%
\ref{M15}) accordingly yield%
\begin{equation}
\text{Tr }(M^{\pm }(a_{0}^{\pm },c_{0}^{\pm })=\pm <c_{(0)}^{\pm }\mid
a_{(0)}^{\pm }>_{g^{\pm }},\text{ Tr }(M^{\pm }(a_{1}^{\pm },c_{1}^{\pm
})=\pm <c_{(1)}^{\pm }\mid a_{(1)}^{\pm }>_{g^{\pm }},  \label{M500}
\end{equation}%
whereas%
\begin{equation}
\text{Tr }M^{\pm }(\mathcal{A}^{\pm },\mathcal{C}^{\pm })=\text{Tr }(M^{\pm
}(a_{0}^{\pm },c_{0}^{\pm })+M^{\pm }(a_{1}^{\pm },c_{1}^{\pm }))=C_{\pm
}^{\sigma }\Delta _{\sigma \mu }^{\pm }\overline{A_{\pm }^{\mu }}.
\label{M19}
\end{equation}%
Equations (\ref{J91}) and \ref{J93} represent a compound apparatus whose
first device prepares a reduced state $<a_{(\mu )}^{\pm }b^{\pm }c^{\pm
}\mid $ from an incident outer product $<a^{\pm }b^{\pm }c^{\pm }\mid ,$ and
allows for it to be transmitted to the second device which, in turn, accepts
a fraction of the prepared $\mathcal{A}^{\pm }$-states and produces a
reduced state $<b_{(\lambda )}^{\pm }\mid .$ The third device then
annihilates the $\mathcal{A}^{\pm }$-states and accepts a fraction of the $%
\mathcal{B}^{\pm }$-states coming from the second stage of the measurement
process, while creating thereafter an outgoing reduced state $<b_{(\lambda
)}^{\pm }c_{(\sigma )}^{\pm }\mid $ from the action of $\Pi _{(\sigma
)}^{\pm }(\mathcal{C}^{\pm }).$ The overall statistical relations that
characterize the accepted and transmitted $\mathcal{B}^{\pm }$-states are
supplied by%
\begin{equation}
<a_{(0)}^{\pm }\mid b_{(0)}^{\pm }>_{g^{\pm }}<b_{(0)}^{\pm }\mid
c_{(0)}^{\pm }>_{g^{\pm }},\text{ }<a_{(1)}^{\pm }\mid b_{(1)}^{\pm
}>_{g^{\pm }}<b_{(1)}^{\pm }\mid c_{(1)}^{\pm }>_{g^{\pm }}.  \label{trans}
\end{equation}%
The intermediate-stage device of (\ref{J15}) may be taken to play a formal
catalytic role during the occurrence of the process if it is replaced with $%
\pi _{(\lambda )}^{\pm }(\mathcal{B}^{\pm }),$ in which case the
completeness of (\ref{J90}) would afford by means of (\ref{M500}) the
following alternative relations for the incoming and outgoing states
produced by the whole process%
\begin{equation}
<a_{(0)}^{\pm }\mid c_{(0)}^{\pm }>_{g^{\pm }},\text{ }<a_{(1)}^{\pm }\mid
c_{(1)}^{\pm }>_{g^{\pm }}.  \label{trans1}
\end{equation}%
Hence, the process as a whole could be described by a device like $M^{\pm
}(a_{\mu }^{\pm },c_{\sigma }^{\pm })$ whose overall operationality
transmits a fraction of a reduced state $<a_{(\mu )}^{\pm }\mid $ and
creates a reduced state $<c_{(\sigma )}^{\pm }\mid .$

As we have seen, the device represented by (\ref{J15}) may transmit
information on a fraction of an incoming reduced state while creating an
outgoing reduced state after eventually hiding a middle-stage state for an
observable that partakes of the process under a dynamical condition of
spectral completeness. A measurement associated to a sequence like%
\begin{equation}
M^{\pm }(a_{\mu }^{\pm },b_{\nu }^{\pm })M^{\pm }(c_{\lambda }^{\pm
},d_{\sigma }^{\pm }),  \label{M51}
\end{equation}%
gives rise to characteristic statistical relationships between incoming and
outgoing reduced states which pertain to the operationality of devices of
the same type as (\ref{d5}). The apparatus symbolized by (\ref{M51}) serves,
in effect, to perform a compound measurement whose first device prepares a
state $<a_{(\mu )}^{\pm }\mid $ and creates a reduced state $<b_{(\nu
)}^{\pm }\mid $ from an incoming state $<a^{\pm }b^{\pm }c^{\pm }d^{\pm
}\mid $, whereas the second device accepts a fraction of $<a_{(\mu )}^{\pm
}b_{(\nu )}^{\pm }\mid $ and produces the reduced state $<c_{(\lambda
)}^{\pm }d^{\pm }\mid $ while ultimately supplying an outgoing reduced state 
$<a_{(\mu )}^{\pm }b_{(\nu )}^{\pm }c_{(\lambda )}^{\pm }d_{(\sigma )}^{\pm
}\mid .$ Then, calling for (\ref{M50}) gives%
\begin{equation}
M^{\pm }(a_{\mu }^{\pm },b_{\nu }^{\pm })M^{\pm }(c_{\lambda }^{\pm
},d_{\sigma }^{\pm })=\pm <b_{(\nu )}^{\pm }\mid c_{(\lambda )}^{\pm
}>_{g^{\pm }}M^{\pm }(a_{\mu }^{\pm },d_{\sigma }^{\pm }),  \label{M52}
\end{equation}%
with $\mu =\nu =\lambda =\sigma .$ It follows that (\ref{M51}) can, in the
last instance, be viewed as a selective measurement operator like $M^{\pm
}(a_{\mu }^{\pm },d_{\sigma }^{\pm }),$ with $<b_{(\nu )}^{\pm }\mid
c_{(\lambda )}^{\pm }>_{g^{\pm }}$ carrying in this way information on an
intermediate transmission amount. Thus, by replacing kernel letters from (%
\ref{M15}), we obtain%
\begin{equation}
<e_{(0)}^{\pm }\mid M^{\pm }(a_{0}^{\pm },b_{0}^{\pm })M^{\pm }(c_{0}^{\pm
},d_{0}^{\pm })\parallel e_{(0)}^{\pm }>_{g^{\pm }}=<b_{(0)}^{\pm }\mid
c_{(0)}^{\pm }>_{g^{\pm }}D_{\pm }^{0}\overline{A_{\pm }^{0}}  \label{J5}
\end{equation}%
and%
\begin{equation}
<e_{(1)}^{\pm }\mid M^{\pm }(a_{1}^{\pm },b_{1}^{\pm })M^{\pm }(c_{1}^{\pm
},d_{1}^{\pm })\parallel e_{(1)}^{\pm }>_{g^{\pm }}=<b_{(1)}^{\pm }\mid
c_{(1)}^{\pm }>_{g^{\pm }}D_{\pm }^{1}\overline{A_{\pm }^{1}},  \label{J6}
\end{equation}%
together with the characteristic statistical trace%
\begin{equation}
\text{Tr }M^{\pm }(a_{\mu }^{\pm },b_{\nu }^{\pm })M^{\pm }(c_{\lambda
}^{\pm },d_{\sigma }^{\pm })=<b_{(\nu )}^{\pm }\mid c_{(\lambda )}^{\pm
}>_{g^{\pm }}<d_{(\sigma )}^{\pm }\mid a_{(\mu )}^{\pm }>_{g^{\pm }},
\label{J100}
\end{equation}%
with $\mu =\nu =\lambda =\sigma $ as before.

A particular couple of similar operators emerges from (\ref{M51}) when we
identify in (\ref{m5}) $\mathcal{A}^{\pm }$ with $\mathcal{B}^{\pm }$ and $%
\mathcal{C}^{\pm }$ with $\mathcal{D}^{\pm },$ while suitably allowing for
in either case some measurement operators coming from (\ref{m6}). Hence, by
accounting for the decomposition (\ref{add102}) for each of $\mathcal{A}%
^{\pm }$ and $\mathcal{C}^{\pm }$ which, by (\ref{M50}), provides Born
constituents like%
\begin{equation}
\Pi _{(0)}^{\pm }(\Sigma ^{\pm })=\mid s_{(0)}^{\pm }>g_{\pm }^{\ast
00}<s_{(0)}^{\pm }\mid =M^{\pm }(s_{0}^{\pm },s_{0}^{\pm })  \label{Add1}
\end{equation}%
and%
\begin{equation}
\Pi _{(1)}^{\pm }(\Sigma ^{\pm })=\mid s_{(1)}^{\pm }>g_{\pm }^{\ast
11}<s_{(1)}^{\pm }\mid =M^{\pm }(s_{1}^{\pm },s_{1}^{\pm }),  \label{Add2}
\end{equation}%
after recalling (\ref{M52}), we arrive at the correspondences%
\begin{equation}
\mathcal{A}^{\pm }\equiv \mathcal{B}^{\pm }\text{: }M^{\pm }(a_{\mu }^{\pm
},b_{\nu }^{\pm })M^{\pm }(c_{\lambda }^{\pm },d_{\sigma }^{\pm
})\longmapsto \Pi _{(\mu )}^{\pm }(\mathcal{A}^{\pm })M^{\pm }(c_{\lambda
}^{\pm },d_{\sigma }^{\pm })  \label{C1}
\end{equation}%
and%
\begin{equation}
\mathcal{C}^{\pm }\equiv \mathcal{D}^{\pm }\text{: }M^{\pm }(a_{\mu }^{\pm
},b_{\nu }^{\pm })M^{\pm }(c_{\lambda }^{\pm },d_{\sigma }^{\pm
})\longmapsto M^{\pm }(a_{\mu }^{\pm },b_{\nu }^{\pm })\Pi _{(\lambda
)}^{\pm }(\mathcal{C}^{\pm }).  \label{C2}
\end{equation}%
Equation (\ref{C1}) thus yields%
\begin{equation}
\text{for }\mu =0\text{: }\Pi _{(0)}^{\pm }(\mathcal{A}^{\pm })M^{\pm
}(c_{0}^{\pm },d_{0}^{\pm })=\pm <a_{(0)}^{\pm }\mid c_{(0)}^{\pm }>_{g^{\pm
}}M^{\pm }(a_{0}^{\pm },d_{0}^{\pm })  \label{C3}
\end{equation}%
and%
\begin{equation}
\text{for }\mu =1\text{: }\Pi _{(1)}^{\pm }(\mathcal{A}^{\pm })M^{\pm
}(c_{1}^{\pm },d_{1}^{\pm })=\pm <a_{(1)}^{\pm }\mid c_{(1)}^{\pm }>_{g^{\pm
}}M^{\pm }(a_{1}^{\pm },d_{1}^{\pm }),  \label{C4}
\end{equation}%
whereas (\ref{C2}) leads to%
\begin{equation}
\text{for }\mu =0\text{: }M^{\pm }(a_{0}^{\pm },b_{0}^{\pm })\Pi _{(0)}^{\pm
}(\mathcal{C}^{\pm })=\pm <b_{(0)}^{\pm }\mid c_{(0)}^{\pm }>_{g^{\pm
}}M^{\pm }(a_{0}^{\pm },c_{0}^{\pm })  \label{C5}
\end{equation}%
and%
\begin{equation}
\text{for }\mu =1\text{: }M^{\pm }(a_{1}^{\pm },b_{1}^{\pm })\Pi _{(1)}^{\pm
}(\mathcal{C}^{\pm })=\pm <b_{(1)}^{\pm }\mid c_{(1)}^{\pm }>_{g^{\pm
}}M^{\pm }(a_{1}^{\pm },c_{1}^{\pm }).  \label{C6}
\end{equation}%
For (\ref{C1}), for instance, we have the values%
\begin{equation}
\text{Tr }(\Pi _{(0)}^{\pm }(\mathcal{A}^{\pm })M^{\pm }(c_{0}^{\pm
},d_{0}^{\pm }))=<d_{(0)}^{\pm }\mid a_{(0)}^{\pm }>_{g^{\pm }}<a_{(0)}^{\pm
}\mid c_{(0)}^{\pm }>_{g^{\pm }}  \label{C7}
\end{equation}%
and%
\begin{equation}
\text{Tr }(\Pi _{(1)}^{\pm }(\mathcal{A}^{\pm })M^{\pm }(c_{1}^{\pm
},d_{1}^{\pm }))=<d_{(1)}^{\pm }\mid a_{(1)}^{\pm }>_{g^{\pm }}<a_{(1)}^{\pm
}\mid c_{(1)}^{\pm }>_{g^{\pm }},  \label{C9}
\end{equation}%
which bring back the statistical shape of (\ref{trans1}) whenever $\Pi
_{(\mu )}^{\pm }(\mathcal{A}^{\pm })$ is replaced by $\pi _{(\mu )}^{\pm }(%
\mathcal{A}^{\pm }).$

\section*{Observational correlations}

In the $g$-realization of $SU(2,2)$, the representation of the orthochronous
proper Poincar\'{e} group consists [13,15,28,29] of the set $\mathcal{P}%
_{+}^{\uparrow }$ of all ten-parameter matrices of the form%
\begin{equation}
U(\mathcal{P}_{+}^{\uparrow }){}=\frac{1}{2}\left( 
\begin{array}{cc}
a+(I_{2}+iW)a^{-1\dagger } & a-(I_{2}+iW)a^{-1\dagger } \\ 
-a+(I_{2}-iW)a^{-1\dagger } & -a-(I_{2}-iW)a^{-1\dagger }%
\end{array}%
\right) .  \label{f1}
\end{equation}%
In Eq. (\ref{f1}), $a$ denotes an element of $SL(2,%
\mathbb{C}
)$ such that it corresponds to an orthochronous proper Lorentz
transformation, and $W$ is the van der Waerden Hermitian $(2\times 2)$%
-matrix associated to a time-like or space-like Minkowskian translation. The
representation of the orthochronous proper Lorentz group is thus constituted
by the set $\mathcal{L}_{+}^{\uparrow }$ of all six-parameter matrices of
the type\footnote{%
Matrix blocks that enter any realizations of $SU(2,2)$ typically bear a
skew-Hermiticity property [24,28,29].}%
\begin{equation}
U{}(\mathcal{L}_{+}^{\uparrow })=\frac{1}{2}\left( 
\begin{array}{cc}
a+a^{-1\dagger } & a-a^{-1\dagger } \\ 
-a+a^{-1\dagger } & -a-a^{-1\dagger }%
\end{array}%
\right) .  \label{frame2}
\end{equation}

The correlations between the physically meaningful information that comes
from quantum measurements performed in different spacetime frames, are taken
to be supplied by the dynamical intersection%
\begin{equation}
SU(4)\supset \mathcal{P}_{Dyn}^{+\uparrow }=\mathcal{P}_{+}^{\uparrow }\cap
U(4).  \label{frame3}
\end{equation}%
Any changes of spacetime frames are induced by the action of $\mathcal{P}%
_{Dyn}^{+\uparrow }$ on some admissible bases, with $\mathcal{P}%
_{Dyn}^{+\uparrow }$ accordingly consisting of the totality of
seven-parameter Poincar\'{e} matrices of the form%
\begin{equation}
U(\mathcal{P}_{Dyn}^{+\uparrow }){}=\frac{1}{\sqrt{2}}\left( 
\begin{array}{cc}
(I_{2}+iW)\beta  & 0_{2} \\ 
0_{2} & \beta ^{\dag }(I_{2}-iW)%
\end{array}%
\right) ,  \label{frame4}
\end{equation}%
where $\beta \in U(2)$ and $W$ has now to be normalized as $W^{2}=I_{2}.$
Dropping $W$ from $U(\mathcal{P}_{Dyn}^{+\uparrow })$ calls for the
suppression of the unimodularization factor from (\ref{frame4}), and thence
produces the four-parameter unitary pattern\footnote{%
The statement $\beta \in SL(2,\mathbf{C})\cap U(2)$ carried by Eq. (7.2) of
Ref. [13] should read $\beta \in U(2)$ whence the representation of the
operator (7.1) lying thereabout must bear seven real parameters instead of
six.}%
\begin{equation}
\left( 
\begin{array}{cc}
\beta  & 0_{2} \\ 
0_{2} & \beta ^{\dag }%
\end{array}%
\right) \in \mathcal{L}_{+}^{\uparrow }\cap \mathcal{P}_{Dyn}^{+\uparrow }.
\label{f5}
\end{equation}%
A restriction represented by an array like (\ref{frame4}) appears as%
\begin{equation}
u(\mathcal{P}_{Dyn}^{+\uparrow }){}=\left( 
\begin{array}{ll}
u_{Dyn}^{+}{} & 0 \\ 
0 & u_{Dyn}^{-}{}%
\end{array}%
\right) ,\text{ }u_{Dyn}^{\pm \bigstar }{}=(u_{Dyn}^{\pm
}{})^{-1}=u_{Dyn}^{\pm \dag }{},  \label{frame5}
\end{equation}%
with the prototype entry%
\begin{equation}
u_{\mu \nu }^{\pm }(Dyn){}=<e_{(\mu )}^{\pm }\mid u_{Dyn}^{\pm }{}\parallel
e_{(\nu )}^{\pm }>_{g^{\pm }},  \label{frame9}
\end{equation}%
which, by (\ref{77}) and (\ref{78}), must satisfy%
\begin{equation}
u_{\mu \lambda }^{\pm }(Dyn){}g_{\pm }^{\ast \lambda \sigma }u_{\sigma \nu
}^{\pm \bigstar }(Dyn){}=g_{\mu \nu }^{\pm }.  \label{f10}
\end{equation}

We assume that the copy of Cartan's space for any spacetime frame admits a
decomposition just like $\mathfrak{C}^{\pm }$ such that Eq. (\ref{11}) must
bear $\mathcal{P}_{Dyn}^{+\uparrow }$-invariance (see (\ref{frame51})
below). Hence, the computational bases for any two frames are subject to a
relation that looks like%
\begin{equation}
<e_{(\mu )}^{\prime \pm }\mid =<e_{(\mu )}^{\pm }\mid u_{Dyn}^{\pm }{},
\label{frame6}
\end{equation}%
whence the behaviours under $\mathcal{P}_{Dyn}^{+\uparrow }$ of both $g_{\mu
\nu }^{\pm }$ and $\Delta _{\mu \nu }^{\pm }$ can be exhibited by%
\begin{equation}
<e_{(\mu )}^{\prime \pm }\mid e_{(\nu )}^{\prime \pm }>_{g^{\prime \pm
}}=g_{\mu \nu }^{\prime \pm }=<e_{(\mu )}^{\pm }\mid u_{Dyn}^{\pm }{}\mid
u_{Dyn}^{\pm }{}\mid e_{(\nu )}^{\pm }>_{g^{\pm }}=g_{\mu \nu }^{\pm }
\label{f2}
\end{equation}%
and%
\begin{equation}
<<e_{(\mu )}^{\prime \pm }\mid e_{(\nu )}^{\prime \pm }>>=\Delta _{\mu \nu
}^{\prime \pm }=<<e_{(\mu )}^{\pm }\mid u_{Dyn}^{\pm }{}\mid u_{Dyn}^{\pm
}{}\mid e_{(\nu )}^{\pm }>>=\Delta _{\mu \nu }^{\pm },  \label{f3}
\end{equation}%
which apply in form equally well to $g_{\pm }^{\ast \mu \nu }$ and $\Delta
_{\pm }^{\ast \mu \nu }$ too. Thus, we also have%
\begin{equation}
g_{\mu \nu }^{\prime \pm }=u_{\mu \lambda }^{\pm }(Dyn)g_{\pm }^{\ast
\lambda \sigma }{}u_{\sigma \nu }^{\pm \bigstar }(Dyn)=g_{\mu \nu }^{\pm }
\label{f12}
\end{equation}%
and%
\begin{equation}
\Delta _{\mu \nu }^{\prime \pm }=u_{\mu \lambda }^{\pm }(Dyn)\Delta _{\pm
}^{\ast \lambda \sigma }{}u_{\sigma \nu }^{\pm \dag }(Dyn)=\Delta _{\mu \nu
}^{\pm },  \label{f15}
\end{equation}%
whilst the behaviour of (\ref{50}) can be specified from%
\begin{equation}
\mid e_{(\mu )}^{\prime \pm }>g_{\pm }^{\prime \ast \mu \nu }<e_{(\nu
)}^{\prime \pm }\mid =\mid e_{(\mu )}^{\pm }>u_{\pm }^{\ast \bigstar \mu
\lambda }(Dyn)g_{\lambda \sigma }^{\pm }u_{\pm }^{\ast \sigma \nu
}(Dyn)<e_{(\nu )}^{\pm }\mid .  \label{f}
\end{equation}%
The pattern for the adjoint version of (\ref{f12}) then yields the
invariance property%
\begin{equation}
\mid e_{(\mu )}^{\prime \pm }>g_{\pm }^{\prime \ast \mu \nu }<e_{(\nu
)}^{\prime \pm }\mid =\mid e_{(\mu )}^{\pm }>g_{\pm }^{\ast \mu \nu
}<e_{(\nu )}^{\pm }\mid .  \label{f16}
\end{equation}

Any state vector like (\ref{115}) is naturally invariant under $\mathcal{P}%
_{Dyn}^{+\uparrow }$. This assertion relies partially upon the fact that the
effective action of $\mathcal{P}_{Dyn}^{+\uparrow }$ requires that%
\begin{equation}
<s^{\prime \pm }\mid s^{\prime \pm }>_{g^{\prime \pm }}=<s^{\pm }\mid s^{\pm
}>_{g^{\pm }},  \label{f17}
\end{equation}%
whence we have to encompass the $\bigstar $-conjugate configurations%
\begin{equation}
<s^{\prime \pm }\mid =S_{\pm }^{\mu }u_{\mu \sigma }^{\pm \bigstar
}{}(Dyn)u_{\pm }^{\ast \sigma \lambda }(Dyn)<e_{(\lambda )}^{\pm }\mid
=<s^{\pm }\mid  \label{f18}
\end{equation}%
and%
\begin{equation}
\mid s^{\prime \pm }>=\mid e_{(\rho )}^{\pm }>u_{\pm }^{\ast \bigstar \rho
\tau }(Dyn)u_{\tau \nu }^{\pm }{}(Dyn)\overline{S_{\pm }^{\nu }}=\mid s^{\pm
}>,  \label{f90}
\end{equation}%
which take up (\ref{55}) as well as the auxiliary relations%
\begin{equation}
u_{\mu }^{\pm \bigstar }{}^{\nu }(Dyn)=g_{\mu \lambda }^{\pm }\overline{%
u^{\pm \lambda }{}_{\sigma }}(Dyn)g_{\pm }^{\ast \sigma \nu }  \label{f91}
\end{equation}%
and\footnote{%
It is shown in Ref. [18] that $g_{\mu \lambda }^{\pm }g_{\pm }^{\ast \lambda
\nu }=\Delta _{\mu \lambda }^{\pm }\Delta _{\pm }^{\ast \lambda \nu }.$ When
combined with (\ref{f12}), this property guarantees the legitimacy of (\ref%
{f18}) and (\ref{f90}).}%
\begin{equation}
u_{\mu }^{\pm \bigstar }{}^{\sigma }(Dyn)u_{\sigma }^{\pm }{}^{\lambda
}(Dyn)=u_{\mu \sigma }^{\pm \bigstar }{}(Dyn)u_{\pm }^{\ast \sigma \lambda
}(Dyn).  \label{f50}
\end{equation}%
Therefore, the defining invariance of $g^{\pm }$ assures that both (\ref{j1}%
) and (\ref{j2}) are invariant, while the normalization condition (\ref%
{add51}) can be appropriately regarded as an invariant dynamical feature
albeit each of the involved amplitudes can not at all. We have, for instance,%
\begin{equation}
S_{\pm }^{\prime \lambda }=S_{\pm }^{\mu }u_{\mu \sigma }^{\pm \bigstar
}{}(Dyn)g_{\pm }^{\ast \sigma \lambda }.  \label{f19}
\end{equation}%
It follows that any reduced states like those defined by (\ref{m10}) do not
hold $\mathcal{P}_{Dyn}^{+\uparrow }$-invariance, but their sum does since (%
\ref{frame6}) and (\ref{f19}) may recover (\ref{f18}), say, from%
\begin{equation}
<s_{(0)}^{\prime \pm }\mid +<s_{(1)}^{\prime \pm }\mid =S_{\pm }^{\mu
}u_{\mu \lambda }^{\pm \bigstar }{}(Dyn)(g_{\pm }^{\ast \lambda 0}g_{0\rho
}^{\pm }+g_{\pm }^{\ast \lambda 1}g_{1\rho }^{\pm })u_{\pm }^{\ast \rho
\sigma }(Dyn)<e_{(\sigma )}^{\pm }\mid ,  \label{F}
\end{equation}%
whence both of (\ref{state1}) and (\ref{m50}) behave invariantly.

The behaviours under $\mathcal{P}_{Dyn}^{+\uparrow }$ of dynamical variables
and spectra depend closely on the physical nature of the magnitudes being
observed. For bringing out the immediately relevant situations, we allow for
the primed-frame entry%
\begin{equation}
S_{\mu \nu }^{\prime \pm }=<e_{(\mu )}^{\prime \pm }\mid S^{\prime \pm
}\parallel e_{(\nu )}^{\prime \pm }>_{g^{\prime \pm }},  \label{f21}
\end{equation}%
to write down the correlations%
\begin{equation}
S^{\prime \pm }=S^{\pm }\Longrightarrow S_{\mu \nu }^{\prime \pm }=u_{\mu
\lambda }^{\pm }{}(Dyn)S_{\pm }^{\ast \lambda \sigma }u_{\sigma \nu }^{\pm
\bigstar }(Dyn)  \label{f32}
\end{equation}%
and\footnote{%
The diagonazibility of diagonal arrays is generally lost when (\ref{f32})
comes about. In either of the situations of (\ref{f32}) and (\ref{f36}), the
compatibility of observables remains invariant.}%
\begin{equation}
S_{\mu \nu }^{\prime \pm }=S_{\mu \nu }^{\pm }\Longrightarrow S^{\prime \pm
}=u_{Dyn}^{\pm \bigstar }{}S^{\pm }u_{Dyn}^{\pm }{}.  \label{f36}
\end{equation}%
The spectrum (\ref{700}) is required to bear $\mathcal{P}_{Dyn}^{+\uparrow }$%
-invariance in accordance with the behaviour specified by (\ref{f36}) such
that the eigenvalues carried by (\ref{m2}), (\ref{m2lin}) and (\ref{109})
are subject to%
\begin{equation}
s_{0}^{\prime \pm }=s_{0}^{\pm },\text{ }s_{1}^{\prime \pm }=s_{1}^{\pm },
\label{f43}
\end{equation}%
whereas Eq. (\ref{300}) undergoes the transformation%
\begin{equation}
<S^{\prime \pm }>_{s^{\prime \pm }}=\pm s_{\mu }^{\pm }\mid S_{\pm
}^{\lambda }u_{\lambda \sigma }^{\pm \bigstar }{}(Dyn)g_{\pm }^{\ast \sigma
\mu }\mid ^{2}.  \label{frame50}
\end{equation}

Formal traces of observables, as given by (\ref{88}) and (\ref{89}), hold $%
\mathcal{P}_{Dyn}^{+\uparrow }$-invariance in the case of either of Eqs. (%
\ref{f32}) and (\ref{f36}). The case of (\ref{f36}) is treated as%
\begin{equation}
\text{Tr }S^{\prime \pm }=\text{Tr }(u_{Dyn}^{\pm \bigstar }{}S^{\pm
}u_{Dyn}^{\pm }{})=\text{Tr }(u_{Dyn}^{\pm }{}u_{Dyn}^{\pm \bigstar
}{}S^{\pm })=\text{Tr }S^{\pm },  \label{T1}
\end{equation}%
and it agrees with (\ref{f43}). Owing to the adjoint version of (\ref{f12}),
use can be made of (\ref{55}) to perform the following calculation for (\ref%
{f32})%
\begin{equation}
S_{\mu \nu }^{\prime \pm }g_{\pm }^{\prime \ast \nu \mu }=u_{\mu \lambda
}^{\pm }g_{\pm }^{\ast \lambda \tau }S_{\tau \sigma }^{\pm }g_{\pm }^{\ast
\sigma \rho }g_{\rho \theta }^{\pm }u_{\pm }^{\ast \bigstar \theta \mu
}=S_{\tau \sigma }^{\pm }g_{\pm }^{\ast \sigma \rho }g_{\rho \theta }^{\pm
}u_{\pm }^{\ast \bigstar \theta \mu }u_{\mu \lambda }^{\pm }g_{\pm }^{\ast
\lambda \tau }=S_{\tau \sigma }^{\pm }g_{\pm }^{\ast \sigma \tau },
\label{t}
\end{equation}%
with the parenthesized $Dyn$ having been suppressed just for once. It should
be obvious that Eqs. (\ref{f32}) and (\ref{f36}) still apply formally to
operators other than observables. In particular, the invariant character of (%
\ref{11}) implies that the representative matrices for the operators $P^{\pm
}$ defined by Eq. (\ref{15}) must bear $\mathcal{P}_{Dyn}^{+\uparrow }$%
-invariance in accordance with (\ref{f36}), namely,%
\begin{equation}
P_{\mu \nu }^{\prime \pm }=P_{\mu \nu }^{\pm }\Longrightarrow P^{\prime \pm
}=u_{Dyn}^{\pm \bigstar }{}P^{\pm }u_{Dyn}^{\pm }{}.  \label{frame51}
\end{equation}

Any measurement processes should run covariantly with respect to the action
of $\mathcal{P}_{Dyn}^{+\uparrow },$ which means that any process must admit
invariant operator patterns. This requirement is certainly fulfilled by the
implementation of Eq. (\ref{f32}) and, therefore, also agrees with the
non-invariant character due to (\ref{f19}) and its complex conjugate of any
representations that carry individual amplitudes alone.

The state reduction produced by the operator $\pi _{(\mu )}^{\pm }(\Sigma
^{\pm })$ that occurs on the right-hand side of Eq. (\ref{m9}), gives rise
to a loss of invariance which may be posed as%
\begin{equation}
\pi _{(\mu )}^{\pm }(\Sigma ^{\pm })\text{: invariant}<s^{\pm }\mid
\longmapsto \text{non-invariant }<s_{(\mu )}^{\pm }\mid ,  \label{fr52}
\end{equation}%
but an invariance recovery may be attained from (\ref{F}) according to%
\begin{equation}
<s^{\pm }\mid (\pi _{(0)}^{\pm }(\Sigma ^{\pm })+\pi _{(1)}^{\pm }(\Sigma
^{\pm }))=\text{invariant,}  \label{f53}
\end{equation}%
which reflects the behaviour of (\ref{J90}), namely,%
\begin{equation}
\mid e_{(\rho )}^{\pm }>g_{\pm }^{\ast \rho \lambda }\pi _{(\mu )\lambda
\sigma }^{\pm }(\Sigma ^{\pm })g_{\pm }^{\ast \sigma \tau }<e_{(\tau )}^{\pm
}\mid =\text{invariant.}  \label{fr90}
\end{equation}%
Hence, the invariance of (\ref{add51}) gets translated into the statement%
\begin{equation}
<s^{\pm }\mid \pi _{(0)}^{\pm }(\Sigma ^{\pm })g^{\pm }\parallel s^{\pm
}>_{g^{\pm }}+<s^{\pm }\mid \pi _{(1)}^{\pm }(\Sigma ^{\pm })g^{\pm
}\parallel s^{\pm }>_{g^{\pm }}=\text{invariant,}  \label{f54}
\end{equation}%
while the representations (\ref{m15}) must behave as%
\begin{equation}
\pi _{(\mu )\lambda \sigma }^{\prime \pm }(\Sigma ^{\prime \pm })=u_{\lambda
\rho }^{\pm }{}(Dyn)\pi _{\pm }^{\ast (\mu )\rho \tau }{}(\Sigma ^{\pm
})u_{\tau \sigma }^{\pm \bigstar }(Dyn).  \label{f55}
\end{equation}%
In addition, the invariance property exhibited by (\ref{t}) applies to (\ref%
{m15}), that is to say,%
\begin{equation}
\text{Tr (}\pi _{(0)}^{\prime \pm }{}(\Sigma ^{\prime \pm })+\pi
_{(1)}^{\prime \pm }{}(\Sigma ^{\prime \pm }))=\text{Tr (}\pi _{(0)}^{\pm
}{}(\Sigma ^{\pm })+\pi _{(1)}^{\pm }{}(\Sigma ^{\pm })).  \label{frame12}
\end{equation}%
When combined together with (\ref{frame6}) and its dual, Eq. (\ref{f55})
shows that the decomposition (\ref{m100}) is invariant.

The $\mathcal{P}_{Dyn}^{+\uparrow }$-behaviour of (\ref{add51}) implies that
Eqs. (\ref{1*}) and (\ref{2*}) should be subjected to%
\begin{equation}
<s_{(0)}^{\pm }\mid \pi ^{\pm }(s_{0}^{\pm },s_{1}^{\pm })\parallel
s_{(1)}^{\pm }>_{g^{\pm }}+<s_{(1)}^{\pm }\mid \pi ^{\pm }(s_{1}^{\pm
},s_{0}^{\pm })\parallel s_{(0)}^{\pm }>_{g^{\pm }}=\text{invariant,}
\label{T}
\end{equation}%
whilst (\ref{f17}) ensures that the relation (\ref{*}) must bear invariance.
Requiring that the operator decomposition (\ref{9*}) should behave
invariantly then yields the following transformation law for the
representations (\ref{3*})%
\begin{equation}
\pi _{\lambda \sigma }^{\prime \pm }(s_{\mu }^{\prime \pm },s_{\nu }^{\prime
\pm })=u_{\lambda \rho }^{\pm }{}(Dyn)g_{\pm }^{\ast \rho \tau }\pi _{\tau
\gamma }^{\pm }(s_{\mu }^{\pm },s_{\nu }^{\pm })g_{\pm }^{\ast \gamma \theta
}u_{\theta \sigma }^{\pm \bigstar }(Dyn).  \label{f57}
\end{equation}

The behaviours of any operator components like (\ref{pi}), (\ref{Pi}) and (%
\ref{d5}) get specified when we bring up the transformation laws for dual
reduced states. By invoking (\ref{F}), we get, for instance, 
\begin{equation}
<s_{(0)}^{\prime \pm }\mid =S_{\pm }^{\mu }u_{\mu \lambda }^{\pm \bigstar
}{}(Dyn)g_{\pm }^{\ast \lambda 0}g_{0\rho }^{\pm }u_{\pm }^{\ast \rho \sigma
}(Dyn)<e_{(\sigma )}^{\pm }\mid  \label{d1}
\end{equation}%
and%
\begin{equation}
\mid s_{(0)}^{\prime \pm }>=\mid e_{(\sigma )}^{\pm }>u_{\pm }^{\ast
\bigstar \sigma \rho }(Dyn)g_{\rho 0}^{\pm }g_{\pm }^{\ast 0\lambda
}u_{\lambda \mu }^{\pm }{}(Dyn)\overline{S_{\pm }^{\mu }},  \label{d2}
\end{equation}%
such that the decompositions (\ref{add102}) and (\ref{M50}) fulfill an
invariant-operator requirement, while (\ref{j2}) and (\ref{M19}) bear
invariance because of the applicability of (\ref{f18}), (\ref{f90}) and (\ref%
{t}). As a consequence, any completion of measurements like the one which
leads to (\ref{J910}) and (\ref{trans1}), is invariant. Likewise, Eqs. (\ref%
{m50}) and (\ref{f17}) together imply that any appropriate sum of
statistical weights is a dynamical invariant, in compatibility with (\ref{t}%
). For (\ref{M500}), for instance, we have%
\begin{equation}
<c_{(0)}^{\pm }\mid a_{(0)}^{\pm }>_{g^{\pm }}+<c_{(1)}^{\pm }\mid
a_{(1)}^{\pm }>_{g^{\pm }}=\text{invariant,}  \label{d3}
\end{equation}%
whence (\ref{M52}) is subjected to%
\begin{equation}
M^{\pm }(a_{0}^{\pm },b_{0}^{\pm })M^{\pm }(c_{0}^{\pm },d_{0}^{\pm
})+M^{\pm }(a_{1}^{\pm },b_{1}^{\pm })M^{\pm }(c_{1}^{\pm },d_{1}^{\pm })=%
\text{invariant,}  \label{d4}
\end{equation}%
and Eqs. (\ref{C3})-(\ref{C6}) thus behave typically as%
\begin{equation}
\Pi _{(0)}^{\pm }(\mathcal{A}^{\pm })M^{\pm }(c_{0}^{\pm },d_{0}^{\pm })+\Pi
_{(1)}^{\pm }(\mathcal{A}^{\pm })M^{\pm }(c_{1}^{\pm },d_{1}^{\pm })=\text{%
invariant.}  \label{d6}
\end{equation}

\section{Concluding remarks and outlook}

The physical role that had been originally assigned to Cartan's space is
reinforced by the extraordinary mathematical fact posed by Segal's theorem
[30] which states that the restricted conformal symmetry of special
relativity is algebraically associated to the maximal spacetime extension of
the Poincar\'{e} group. The combination of this fact with the
Pauli-Weisskopf theorem [31], according to which the occurrence in Nature of
particles and antiparticles does not depend upon any spin values,
considerably strengthens our belief that the twofold framework proposed in
Ref. [13] can supply not only a self-consistent unified mechanical
description of particles and antiparticles, but also arrangements of
relativistic observational correlations between dynamical variables and
measurements of states for fermions and bosons.

Apart from the unitary representations of the Galilei group [32-34], which
are generally infinite-dimensional, the only observational correlations that
can be sorted out by the standard non-relativistic quantum mechanical
framework just emanate from Pauli's $SU(2)$-spin theory [35]. Therefore,
such correlations bear strictly a combination of locality with a spin
character, whence any preparations and measurements of spin one-half states
that are eventually performed locally by an observer can at most be
systematically manipulated by the same observer by carrying out $SO(3)$%
-transformations. A noteworthy implication of this result is that the
underlying operationality of ordinary quantum computational processes makes
it inviable to provide any observational correlations between outcomes of
measurements carried through in spatially separated frames. So, one of the
questions that might arise here is whether the (physical) character of
quantum computation [36] should bear Poincar\'{e} invariance or not if
operational computing tasks were conducted within the realm of special
relativity.

It has become evident from our work that it is mainly the performance of
complete measurements which may hold a manifestly invariant character in
spacetime. Hence, with respect to this, we can say that relativistic quantum
measurement processes should account for an adequate definition of total
measurements regardless of the spin values of the systems to be measured. We
have claimed that the formal mechanisms involved in the traditional
measurement processes may be neatly fitted in with the framework of Ref.
[13], whence the covariant symbolism of Sections 5 and 6 should accordingly
yield the feasibleness of elegantly supplying a refined description of
Wu-Shaknov entangled states of particle-antiparticle pairs [37-41]. From our
perspective, observational correlations must be necessarily incorporated
into the construction of any quantum computing operations that could
explicitly allow for outcomes of measurement devices set upon different
spacetime frames [7].

As we had presupposed before, copies of Cartan's space constitute the only
known spaces of state vectors in which a legitimate relativistic quantum
information theory may be developed side by side with a covariant
measurement-based quantum computation approach. In this connection, all the
standard computational gates can be reconstructed covariantly by defining
suitable gate operators in $\mathfrak{C}^{\pm }$ and utilizing the
representation theory developed in Ref. [13] along with the dynamical
behaviours described by Eqs. (\ref{f32}) and (\ref{f36}). Since practical
entangled states naively carry a reduced character, their behaviours under
arbitrary spacetime displacements must be controlled by $W$-matrices like
the one involved in Eq. (\ref{frame4}). We can then expect that the popular
non-locality property of non-relativistic quantum mechanics should be
thoroughly revisited when any plausible relativistic framework is actually
taken into account. These remarks will probably be entertained elsewhere in
a forthcoming paper.

Data Availability Statement: the data on the article at issue are entirely
available in the repository arXiv on the location
https://arxiv.org/abs/2505.06559

Ethics declaration: not applicable.

Funding Declaration: there was no Funding as regards the elaboration of the
work.

\end{document}